\documentclass[12pt]{article}

\usepackage{epsf,epsfig}
\usepackage{graphics}

\usepackage{cite}
\usepackage{amsmath,amssymb}
\def\be{\begin{equation}}
\def\ee{\end{equation}}
\def\beq{\begin{eqnarray}}
\def\eeq{\end{eqnarray}}

\def\eps{\epsilon}

\def\DB0{\partial B_0}

\def\LL{{\cal L}}

\def\Cl2{\mbox{Cl}_2}
\def\eps{\epsilon}

\def\cO#1{{\cal O}\left( {#1} \right)}

\def\re{\mathrm{Re}}
\def\im{\mathrm{Im}}

\newcommand{\ket}[1]{|#1 \rangle}
\newcommand{\braket}[1]{\langle #1 \rangle}
\newcommand{\Tprod}[1]{{\mathrm T}\lbrack #1 \rbrack}

\def\eps{\epsilon}

\begin{document}
\thispagestyle{empty}

\begin{flushright}
{\small
PITHA~07/05\\
IPPP/07/35\\
CERN-PH-TH-07-107\\
0707.0773 [hep-ph]\\
July 5, 2007}
\end{flushright}

\vspace{\baselineskip}

\begin{center}
\vspace{0.5\baselineskip}
\textbf{\Large\boldmath
Four-fermion production near the $W$ pair \\[0.2cm] production
threshold}\\
\vspace{3\baselineskip}
{\sc M.~Beneke$^a$, P.~Falgari$^a$, C.~Schwinn$^a$, A.~Signer$^b$ and
G.~Zanderighi$^c$}\\
\vspace{0.7cm}
{\sl ${}^a$Institut f\"ur Theoretische Physik E, RWTH Aachen,\\
D--52056 Aachen, Germany\\
\vspace{0.3cm}
${}^b$IPPP, Department of Physics, University of Durham, \\
Durham DH1 3LE, England\\
\vspace{0.3cm}
${}^c$ CERN, 1211 Geneva 23, Switzerland }
\vspace{3\baselineskip}

\vspace*{0.2cm}
\textbf{Abstract}\\
\vspace{1\baselineskip}
\parbox{0.9\textwidth}{
We perform a dedicated study of the four-fermion production
process  $e^{-} e^{+} \rightarrow  \mu^{-} \bar{\nu}_{\mu} u
\bar{d}\,X$ near the $W$ pair-production threshold in view
of the importance of this process for a precise measurement
of the $W$ boson mass. Accurate theoretical predictions for this process
require a systematic treatment of finite-width effects.
We use unstable-particle effective field theory (EFT) to perform an
expansion in the coupling constants, $\Gamma_W/M_W$, and the
non-relativistic velocity $v$ of the $W$ boson up to next-to-leading
order in $\Gamma_W/M_W\sim
\alpha_{ew}\sim v^2$. We find that the dominant theoretical
uncertainty in $M_W$ is currently due to an incomplete treatment
of initial-state radiation. The remaining uncertainty of the
NLO EFT calculation translates into
$\delta M_W\approx$ 10~--~15 MeV, and to about $5$~MeV
with additional input from the NLO four-fermion calculation
in the full theory.}
\end{center}

\newpage
\setcounter{page}{1}

\newpage
\allowdisplaybreaks[2]

\section{Introduction}

The mass of the $W$ gauge boson is a key observable in the search for
virtual-particle effects through electroweak precision measurements.
Its current value, $\hat M_W=(80.403\pm 0.029)\,$GeV~\cite{Yao:2006px},
is determined from a combination of continuum $W$ pair-production
at LEPII and single-$W$ production at the Tevatron.\footnote{This
value refers to the definition of the $W$ mass from a Breit-Wigner
parameterization with a running width as it is adopted in
the experimental analyses. It is related to the pole mass $M_W$
used in this paper by~\cite{Sirlin:1991fd} $\hat M_W-M_W = \Gamma_W^2/(2 M_W)
+ O(\alpha_{ew}^3)$.}
Further measurements of single-$W$ production at the LHC should
reduce the error by a factor of two. Beyond LHC it has been estimated
that an error of $6\,$MeV could be achieved by operating an
$e^- e^+$ collider in the vicinity of the $W$ pair-production
threshold~\cite{Wilson:2001aw}. This estimate is based on
statistics and the performance of a future linear collider, and it
assumes that the cross section is known theoretically to sufficient
accuracy so that its measurement can be converted into one
of $M_W$. In reality, achieving this accuracy is a difficult
theoretical task, requiring the calculation of loop and radiative
corrections. Since the $W$ bosons decay rapidly, this calculation
should be done for a final state of sufficiently long-lived particles,
rather than for on-shell $W$ pair-production. A systematic treatment
of finite-width effects is therefore needed.

In this paper we investigate in detail the
inclusive four-fermion production process
\begin{equation}
e^-(p_1)\, e^+(p_2) \to
\mu^-\, \bar{\nu}_\mu\, u\, \bar{d} + X
\label{eq:wwprocess}
\end{equation}
in the vicinity of the $W$ pair-production threshold, i.e. for
$s\equiv (p_1+p_2)^2\sim 4M_W^2$. Here $X$ denotes an arbitrary
flavour-singlet state (nothing, photons, gluons, ...).
No kinematic cuts shall be applied
to the final state. In this kinematical regime the
process~\eqref{eq:wwprocess} is primarily mediated by the
production of two resonant, non-relativistic $W$ bosons
with virtuality of order
\begin{equation}
  k^2-M_W^2 \sim M_W^2 v^2
\sim M_W\Gamma_W \ll M_W^2,
\end{equation}
one of which decays into leptons, the other into hadrons. Here
we have introduced the non-relativistic velocity $v$, and the $W$ decay width
$\Gamma_W$. We perform a systematic expansion of the total cross section
in the small quantities
\begin{equation}
\alpha_{ew}, \quad \frac{s-4 M_W^2}{4 M_W^2}
\sim  v^2, \quad \frac{\Gamma_W}{M_W}  \sim   \alpha_{ew},
\label{eq:expparameters}
\end{equation}
corresponding to a (re-organized) loop expansion and a kinematic
expansion. All three expansion parameters are of the same order,
and for power-counting purposes we denote them collectively as
$\delta$. Our calculation is accurate at next-to-leading order
(NLO). Note that resonant processes such as~\eqref{eq:wwprocess}
are complicated by the need to account for the width of the
intermediate unstable particles to avoid kinematic singularities
in their propagators. The expansion in the electroweak coupling
$\alpha_{ew}=\alpha/s_w^2$ is therefore not a standard loop
expansion. ($\alpha$ denotes the electromagnetic coupling, and
$s_w^2 \equiv \sin^2\theta_w$ with $\theta_w$ the Weinberg angle.)

NLO calculations of four-fermion production have been done already
some time ago in the continuum (not near threshold) in the double-pole
approximation for the two $W$
propagators~\cite{Beenakker:1998gr,Denner:1999kn,Denner:2000bj} or
with further
simplifications~\cite{Jadach:2000kw,Jadach:2001uu}.
This approximation was supposed to break down for kinematic reasons in
the threshold region. Thus, when this project was begun
\cite{Beneke:2004xd}, there existed only LO calculations in the
threshold region as well as studies of the effect of Coulomb photon
exchanges~\cite{Fadin:1993kg,Fadin:1995fp}, rendering the effective
field theory approach
\cite{Beneke:2003xh,Beneke:2004km,Chapovsky:2001zt} the method of
choice for the NLO calculation. Meanwhile a full NLO calculation of
four-fermion production has been performed in the complex mass
scheme~\cite{Denner:2005es,Denner:2005fg} without any kinematic
approximations, and for the fully differential cross sections in the
continuum or near threshold.  This is a difficult calculation that
required new methods for the numerical evaluation of one-loop
six-point tensor integrals. In comparison, our approach is
computationally simple, resulting in an almost analytic representation
of the result. The drawback is that our approach is not easily
extended to differential cross sections. Nevertheless, we believe that
a completely independent calculation of NLO four-fermion production is
useful, and we shall compare our result to~\cite{Denner:2005es} in
some detail. Having a compact analytic result at hand is also useful
for an investigation of theoretical uncertainties.  Note that while
the full four-fermion NLO calculation~\cite{Denner:2005es,Denner:2005fg}
is a priori of the same accuracy in $\Gamma_W/M_W$ as the NLO
effective-theory result, it includes a subset of higher order terms in
the EFT expansion. We discuss the relevance of these higher order
terms at the end of this paper.

The organization of the paper is as follows. In
Section~\ref{sec:method} we explain our method of calculation. We
focus on  aspects of unstable-particle effective theory that are
specific to pair production near threshold and refer
to~\cite{Beneke:2004km} for those, which are in complete analogy
with the line-shape calculation of a single resonance. The section
ends with a list of all terms that contribute to the NLO result.
We construct the effective-theory expansion of the tree
approximation to the four-fermion cross section in
Section~\ref{sec:tree}. Of course, this calculation can be done
nearly automatically without any expansions with programs such as
Whizard~\cite{Kilian:2001qz},
CompHep~\cite{Pukhov:1999gg,Boos:2004kh} or
MadEvent~\cite{Stelzer:1994ta,Maltoni:2002qb}. The purpose of this
section is to demonstrate the convergence of the expansion towards
the ``exact'' tree-level result, and to provide analytic
expressions for those terms that form part of the NLO calculation
near threshold. In Section~\ref{sec:radcor} we calculate the
radiative corrections required at NLO. These consist of hard loop
corrections to $W$ pair-production and $W$ decay, of Coulomb
corrections up to two photon exchanges, and soft-photon
corrections. The entire calculation is done setting the light
fermion masses to zero, which is a good approximation except for
the initial-state electrons, whose mass is relevant, since the
cross section is not infrared-safe otherwise. In
Section~\ref{sec:isr} we describe how to transform from the
massless, ``partonic'' cross section to the physical cross section
with finite electron mass, including a resummation of large
logarithms $\ln(s/m_e^2)$ from initial-state radiation. Assembling
the different pieces we obtain the full inclusive NLO four-fermion
cross section in terms of compact analytic and numerical
expressions. In Section~\ref{sec:results} we perform a numerical
evaluation of the NLO cross section, estimate the final accuracy,
and compare our result to~\cite{Denner:2005es}, obtaining very
good agreement. We find  that the dominant theoretical uncertainty
in $M_W$ is currently due to an incomplete treatment of
initial-state radiation. The remaining uncertainty of the NLO EFT
calculation translates into $\delta M_W\approx$ 10 -- 15 MeV, and
to about $5\,$MeV with additional input from the NLO four-fermion
calculation in the full theory. We conclude in
Section~\ref{sec:conclude}. Some of the lengthier equations are
separated from the main text and provided in
Appendices~\ref{ap:4ferm} and~\ref{ap:hard1loop}.

\section{Method of calculation}
\label{sec:method}

We extract the inclusive cross section of the process~\eqref{eq:wwprocess}
from the appropriate cuts of the $e^- e^+$ forward-scattering amplitude. For
inclusive observables, where one integrates over the virtualities of the
intermediate resonances, the propagator singularity poses no difficulty, if
the integration contours can be deformed sufficiently far away from the
singularity. This is not possible, however, for the calculation of the
line-shape of a single resonance, and for pair production near threshold (the
pair production equivalent of the resonance region), where the kinematics does
not allow this deformation. The width of the resonance becomes a relevant
scale, and it may be useful to separate the dynamics at this scale from the
dynamics of the short distance fluctuations at the scale of the resonance mass
by constructing an effective field theory.

\subsection{Unstable-particle effective theory for pair production
near threshold}

The following formalism resembles rather closely the formalism
described in \cite{Beneke:2003xh,Beneke:2004km}. The
generalization from a scalar to a vector boson resonance is
straightforward. The pair-production threshold kinematics implies
a change in power counting that is analogous to the difference
between heavy-quark effective theory and non-relativistic QCD.

In $W$ pair-production the short-distance fluctuations are
given by hard modes, whose momentum components are all
of order $M_W$. After integrating out the hard modes, the
forward-scattering amplitude is given by~\cite{Beneke:2004km}
\begin{eqnarray}
\label{eq:master}
&& i {\cal A} =\sum_{k,l} \int d^4 x \,
\braket{e^- e^+ |
\Tprod{i {\cal O}_p^{(k)\dagger}(0)\,i{\cal O}_p^{(l)}(x)}|e^- e^+}
+ \sum_{k} \,\braket{e^- e^+|i {\cal O}_{4e}^{(k)}(0)|e^- e^+}.
\\[-0.5cm]
\nonumber
\end{eqnarray}
The operators ${\cal O}_p^{(l)}(x)$ (${\cal O}_p^{(k)\dagger}(x)$)
in the first term on the right-hand side produce (destroy)
a pair of non-relativistic $W$ bosons. The
second term accounts for the remaining non-resonant contributions.
The matrix elements are to be computed with the effective Lagrangian
discussed below and the operators include
short-distance coefficients due to the hard fluctuations.
Note that there is no separate term for production of
one resonant and one off-shell $W$, since for such configurations
the integrations are not trapped near the singularity of the $W$
propagator. These configurations are effectively short-distance
and included in the non-resonant production-decay operators
${\cal O}_{4e}^{(k)}(0)$.

The effective Lagrangian describes the propagation and interactions
of two non-relativistic, spin-1 fields $\Omega_\pm^i$ representing
the nearly on-shell (potential) $W^\pm$ modes; two sets of collinear fields
for the incoming electron and positron, respectively; and potential
and collinear photon fields. The corresponding momentum scalings
in the center-of-mass frame are:
\begin{equation}
\begin{aligned}
  \text{potential } (p) &: \quad k_0\sim M_W \delta,\,\,
  |\vec{k}|\sim M_W \sqrt{\delta}\\
  \text{soft } (s)&:  \quad k_0\sim |\vec{k}|\sim  M_W \delta\\
  \text{collinear } (c) &:   \quad k_0 \sim M_W\,,\, k^2\sim M_W^2\delta.
\end{aligned}
\end{equation}
The small parameter $\delta$ is either the non-relativistic
velocity squared, $v^2$, related to $(s-4 M_W^2)/(4 M_W^2)$,
or $\Gamma_W/M_W \sim \alpha_{ew}$, since the
characteristic virtuality is never parametrically smaller than
$M_W\Gamma_W$ for an unstable $W$. The interactions of the
collinear modes are given by soft-collinear effective
theory~\cite{Bauer:2000yr,Bauer:2001yt,Beneke:2002ph}. There is
nothing specifically new related to collinear modes in pair
production, and we refer to~\cite{Beneke:2004km} for further
details. As far as the next-to-leading order calculation is
concerned, the soft-collinear Lagrangian allows us to perform
the standard eikonal approximation for the interaction of soft
photons with the energetic electron (positron) in the soft
one-loop correction.

The Lagrangian for the resonance fields is given by the
non-relativistic Lagrangian, generalized to account for
the instability \cite{Beneke:2004xd,Hoang:2004tg}. The terms
relevant at NLO are
\begin{equation}
{\cal L}_{\rm NRQED} = \sum_{a=\mp} \left[\Omega_a^{\dagger i} \left(
i D^0 + \frac{\vec{D}^2}{2 {M}_W} - \frac{\Delta}{2} \right)
\Omega_a^i
+  \Omega_a^{\dagger i}\,
\frac{(\vec{D}^2-M_W \Delta)^2}{8 M_W^3}\,
\Omega_a^i\right].
\label{LNR}
\end{equation}
Here $\Omega_+^i$ and $\Omega_-^i$ ($i=1,2,3$) are non-relativistic, spin-1
destruction fields for particles with electric charge $\pm 1$,
respectively. The interactions with photons is incorporated
through the covariant derivative $D_\mu \Omega_\pm^i \equiv
(\partial_\mu\mp i e A_\mu) \Omega_\pm^i$. The effective theory
does not contain fields for the other heavy particles in the
Standard Model, the $Z$ and Higgs bosons, and the top quark. Their
propagators are always off-shell by amounts of order $M_W^2$ and
therefore their effect is encoded in the short-distance matching
coefficients. In a general $R_\xi$-gauge this also applies to the
pseudo-Goldstone (unphysical Higgs) fields, except in
't~Hooft-Feynman gauge $\xi=1$, where the scalar $W$ and
unphysical charged pseudo-Goldstone modes have masses $M_W$ and
can also be resonant. However, the two degrees of freedom cancel
each other, leading to the same Lagrangian (\ref{LNR}) describing
the three polarization states of a massive spin-1 particle. The
effective Lagrangian has only a U(1) electromagnetic gauge
symmetry as should be expected at scales far below $M_W$. However,
since the short-distance coefficients of the Lagrangian and all
other operators are determined by fixed-order matching of on-shell
matrix elements to the full Standard Model, they are independent
of the gauge parameter in $R_\xi$-gauge by construction. The often
quoted gauge-invariance problems in the treatment of unstable
particles arise only if one performs resummations of perturbation
theory in gauge-dependent quantities such as propagators.

The matching coefficient $\Delta$ in (\ref{LNR}) is obtained
from the on-shell two-point function of a transverse $W$ boson.
``On-shell'' here refers to the complex pole determined from
\begin{equation}
\bar s-\hat M_W^2-\Pi^{W}_T(\bar s)=0
\end{equation}
with $\hat M_W$ any renormalized mass parameter, and $\Pi^{W}_T(q^2)$
the renormalized, transverse self-energy. The solution to this
equation,
\begin{equation}
\label{eq:pole}
\bar s \equiv M_W^2-i M_W\Gamma_W,
\end{equation}
defines the pole mass and the pole width of the $W$. The matching
coefficient is then given by
\begin{equation}
  \Delta \equiv \frac{\bar s-\hat M_W^2}{\hat M_W} \,\,
\stackrel{\rm pole \,\,scheme}{=}\,\, -i\Gamma_W.
\label{eq:Deltadef}
\end{equation}
In the remainder of the paper, we adopt a renormalization convention
where $\hat M_W$ is the pole mass $M_W$, in which case
$\Delta$ is purely imaginary. With $D^0\sim M_W\delta$,
$\vec{D}^2\sim M_W^2 \delta$, and $\Delta\sim M_W\delta$,
we see that the first bilinear term in (\ref{LNR}) consists
of leading-order operators, while the second is suppressed
by one factor of $\delta$, and can be regarded as a
perturbation. Accordingly, the propagator of the $\Omega_\pm$
fields is
\begin{equation}
\frac{i\, \delta^{ij}}{k^0 - \frac{\vec{k}^2}{2 M_W}-\frac{\Delta}{2}}.
\label{OmegaProp}
\end{equation}
The effective theory naturally
leads to a fixed-width form of the resonance propagator.
Note that it would be sufficient to keep only the one-loop expression
for $\Delta$ in the propagator, and to include higher-order
corrections perturbatively.

Loop diagrams calculated using the Lagrangian~\eqref{LNR} receive
contributions from soft and potential photons.\footnote{What we
call ``soft'' here, is usually termed ``ultrasoft'' in the
literature on non-relativistic QCD. There are further modes
(called ``soft'' there) with momentum
$k\sim M_W\sqrt{\delta}$~\cite{Beneke:1997zp}.
In the present context these modes cause, for instance,
a small modification of the QED Coulomb potential due to the one-loop photon
self-energy, but these effects are beyond NLO.}
Since the potential photons do not correspond to
on-shell particles, they can be integrated out, resulting in a non-local
(Coulomb) potential, analogous to potential non-relativistic
QED~\cite{Pineda:1998kn}. Up to NLO the required PNRQED Lagrangian
is
\begin{equation}
\begin{aligned}
{\cal L}_{\text{PNRQED}} &= \sum_{a=\mp} \left[\Omega_a^{\dagger i} \left(
i D_s^0 + \frac{\vec{\partial}^2}{2 {M}_W} - \frac{\Delta}{2} \right)
\Omega_a^i
+  \Omega_a^{\dagger i}\,
\frac{(\vec{\partial }^2-{M}_W \Delta)^2}{8 {M}_W^3}\,
\Omega_a^i\right]\\
&+\int d^3 \vec{r}\, 
\left[\Omega_-^{\dagger i} \Omega^i_-\right]\!(x+\vec r\,)
\left(-\frac{\alpha}{r}\right)
\left [\Omega_+^{\dagger j}\Omega^j_+\right]\!(x).
\label{LPNR}
\end{aligned}
\end{equation}
Only the (multipole-expanded) soft photon $A^0_s(t,0)$ appears in
the covariant derivative $D_s^0$. The potential $W$ field has
support in a region $\sim \delta^{-1}$ in the time direction and
in a region $\sim \delta^{-1/2}$ in each space direction, hence
the measure $d^4 x$ in the action scales as $\delta^{-5/2}$.
Together with $\partial_0\sim \delta$ we find from the kinetic
term that $\Omega_\mp^i \sim \delta^{3/4}$. Analogously we find
that the non-local Coulomb potential scales as
$\alpha/\sqrt{\delta} \sim \alpha/v$.  Since we count $\alpha\sim
v^2$, the Coulomb potential is suppressed by $v$, or
$\alpha^{1/2}$, and need not be resummed, in contrast to the
case of top-quark pair-production near threshold. However, with
this counting the Coulomb enhancement introduces an expansion in
half-integer powers of the electromagnetic coupling, the one-loop
Coulomb correction being a ``N$^{1/2}$LO'' term.

\subsection{Production vertex, production-decay vertices
and the lead\-ing-order cross section} \label{sec:production}

We now turn to the production and production-decay operators
appearing in the representation (\ref{eq:master}) of the
forward-scattering amplitude. The lowest-dimension production
operator must have field content $\left(\bar{e}_{c_2}e_{c_1} \right)
(\Omega_-^{\dagger i} \Omega_+^{\dagger j})$, where the subscripts
on the electron fields stand for the two different direction labels
of the collinear fields. The short-distance coefficients follow
from matching the expansion of the renormalized on-shell matrix
elements for $e^-e^+\to W^- W^+$ in the small relative $W$ momentum
to the desired order in ordinary weak-coupling perturbation
theory.  The on-shell condition for the $W$ lines
implies that their momentum satisfies $k_{1}^2=k_2^2=\bar s =
M_W^2+ M_W\Delta$, but in a perturbative matching calculation this
condition must be fulfilled only to the appropriate order in $\alpha$
and $\delta$.  On the effective-theory side of the matching
equation one also has to add a factor $\sqrt{2 M_W}\,\varpi^{-1/2}$ with
\begin{equation}
\varpi^{-1} \equiv
\left(1+\frac{M_W\Delta+\vec{k}^{\,2}}{M_W^2}
\right)^{\!1/2}
\label{eq:Phinorm}
\end{equation}
for each external $\Omega$ line \cite{Beneke:2004km}.\footnote{This
is the well-known $(E/M)^{1/2}$ factor, which accounts for the
normalization of non-relativistic fields, generalized to
unstable particles and general mass renormalization conventions.}
At tree-level, and at leading order in $\delta$, $\varpi^{-1}=1$.

\begin{figure}[t]
  \begin{center}
  \includegraphics[width=0.6\textwidth]{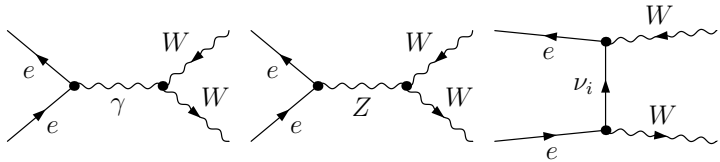}
  \caption{Diagrams contributing to the tree-level matching
  of ${\cal O}_p^{(0)}$.}
  \label{fig:tree}
  \end{center}
\end{figure}

Thus we are led to consider the tree-level, on-shell
$W$ pair-production amplitude shown in Figure~\ref{fig:tree}. To
leading order in the non-relativistic expansion the $s$-channel diagrams
vanish and only the helicity configuration $e^-_L e^+_R$ contributes.
The corresponding operator (including its tree-level coefficient function)
reads
\begin{equation}
{\cal O}_p^{(0)} = \frac{\pi\alpha_{ew}}{M_W^2}
\left(\bar{e}_{c_2,L} \gamma^{[i} n^{j]} e_{c_1,L} \right)
\left(\Omega_-^{\dagger i} \Omega_+^{\dagger j}\right) ,
\label{LPlead}
\end{equation}
where we have introduced the notation
$a^{[i} b^{j]}\equiv a^i b^j + a^j b^i$ and the unit-vector
$\vec{n}$ for the direction of the incoming electron
three-momentum $\vec{p}_1$.
For completeness we note that the emission of collinear photons
from the $W$ or collinear fields of some other direction, which leads
to off-shell propagators, can be incorporated by adding Wilson lines
to the collinear fields, implying the form
$(\bar{e}_{c_2,L}W_{c_2} \gamma^{[i} n^{j]} W_{c_1}^\dagger e_{c_1,L}
)$. However, these Wilson lines will not be needed for our
NLO calculation, since the collinear loop integrals vanish (see,
however, Section~\ref{sec:isr}).

\begin{figure}[t]
  \begin{center}
  \includegraphics[width=0.18\textwidth]{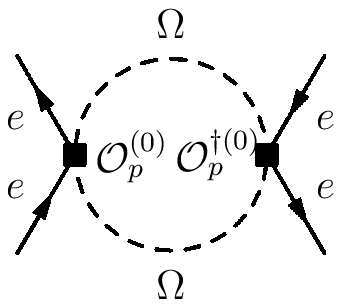}
\caption{Leading-order effective-theory diagram for the
forward-scattering
 amplitude.}
\label{fig:LO}
  \end{center}
\end{figure}

The leading contribution from the potential region to the
forward-scattering amplitude is given by the expression
\begin{equation}
i {\cal A}^{(0)}_{LR}=\int d^4 x\,
\braket{e^-_L e^+_R |\Tprod{i {\cal O}_p^{(0)\dagger }(0)i
{\cal O}_p^{(0)}(x)}|e^-_L e^+_R }.
\end{equation}
This corresponds to the one-loop diagram shown in Figure~\ref{fig:LO},
computed with the vertex~\eqref{LPlead} and the
propagator~\eqref{OmegaProp}. We can use power counting to estimate
the magnitude of the leading-order amplitude prior to its
calculation. With $e_{c_i,L}\sim \delta^{1/2}$, $\Omega^i_\mp
\sim \delta^{3/4}$ the production operator scales as
${\cal O}^{(0)}_p \sim \alpha \delta^{5/2}$.
The integration measure scales as $\int d^4 x\sim
\delta^{-5/2}$ in the potential region and the external
collinear states are normalized as $|e^\mp\rangle \sim \delta^{-1/2}$,
hence ${\cal A}^{(0)}_{LR}\sim \alpha^2\delta^{1/2}$.
This expectation is confirmed by the explicit calculation of the
one-loop diagram:
\begin{eqnarray}
i {\cal A}^{(0)}_{LR} &=&\frac{\pi^2\alpha_{ew}^2 }{M_W^4}\,
\langle p_2-|n^{[i} \gamma^{j]} |p_1-\rangle
\langle p_1-|n^{[i} \gamma^{j]} |p_2-\rangle
\nonumber\\
&&\times \int\frac{d^d r}{(2\pi)^d}
\frac{1}{
\left(r^0 - \frac{\vec{r}^{\,2}}{2 M_W} - \frac{\Delta}{2}\right)
\left(E-r^0 - \frac{\vec{r}^{\,2}}{2M_W} - \frac{\Delta}{2}\right)}
\nonumber \\
 &=&- 4i \pi \alpha_{ew}^2 \,\sqrt{-\frac{E+i \Gamma_W}{M_W}}.
\label{eq:Alead}
\end{eqnarray}
Here we have defined $E=\sqrt s -2 M_W$.
We adopted the standard helicity notation
$\ket{p\pm}=\frac{1\pm\gamma^5}{2}u(p)$, and used
$\Delta =-i \Gamma_W$, valid in the pole scheme, in the last line.
The fermion energies are set to $M_W$ in the external spinors.
The calculation has been
performed by first evaluating the $r^0$ integral using Cauchy's
theorem, and the trace
$\langle p_2-|n^{[i} \gamma^{j]} |p_1-\rangle
\langle p_1-|n^{[i} \gamma^{j]} |p_2-\rangle =
16 (1-\epsilon) M_W^2 $.
The remaining $|\vec r\,|$ integral contains a linear
divergence that is, however, rendered finite by dimensional
regularization (with $d=4-2\epsilon$)
so the $d\to 4$ limit can be taken.
The numerical comparison of~\eqref{eq:Alead} to the
full tree-level result and the convergence of the
effective-theory approximation will be  discussed in
Section~\ref{sec:tree}.

Taking the imaginary part of (\ref{eq:Alead}) does not yield the
cross section of the four-fermion production
process~\eqref{eq:wwprocess} with its flavour-specific final
state. At leading order the correct result is given by
multiplying the imaginary part with the leading-order branching
fraction product $
 \text{Br}^{(0)}(W^-\to \mu^- \bar{\nu}_\mu) \text{Br}^{(0)}(W^+ \to u
 \bar{d}\,)=1/27$.  This procedure can be justified as follows.
The imaginary part of the non-relativistic propagator obtained by
cutting an $\Omega$ line is given by
\begin{equation}
\label{eq:im-propagator}
\im\,\frac{1}{E- \frac{\vec{k}^{\,2}}{2 M_W} + \frac{i \Gamma_W^{(0)}}{2}}
=-\frac{\Gamma_W^{(0)}/2}{
\left(E- \frac{\vec{k}^{\,2}}{2 M_W} \right)^2+\frac{\Gamma_W^{(0)2}}{4}}.
\end{equation}
The propagator of the $\Omega_\pm$ line implicitly includes a string
of self-energy insertions.  Taking the imaginary part amounts to
performing all possible cuts of the self-energy insertions
while the unstable particle is not cut~\cite{Veltman:1963th}. To
obtain the total cross section for a flavour-specific four-fermion
final state, only the cuts through these specific fermion lines
have to be taken into account.
At the leading order this amounts to replacing $\Gamma_W^{(0)}$
in the numerator of (\ref{eq:im-propagator})
by the corresponding partial width, here $\Gamma^{(0)}_{\mu^-\bar\nu_\mu}$
and $\Gamma^{(0)}_{u\bar d}$, respectively,
while the total width is retained in the denominator. The leading-order
cross section is therefore
\begin{equation}
\sigma^{(0)}_{LR} = \frac{1}{27 s}\,\mbox{Im}\,{\cal A}^{(0)}_{LR}
= \frac{4\pi\alpha^2}{27 s_w^4 s} \,\mbox{Im}\left[- \sqrt{-\frac{E+i
\Gamma_W^{(0)}}{M_W}}\,\,\right].
\label{eftLOsigma}
\end{equation}
The unpolarized cross section is given by $\sigma^{(0)}_{LR}/4$,
since the other three helicity combinations vanish.

\begin{figure}[t]
  \begin{center}
  \includegraphics[width=0.9\textwidth]{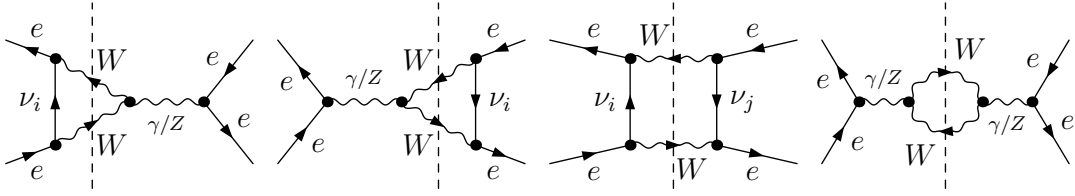}
\caption{Cut one-loop diagrams contributing to non-resonant
  production-decay operator matching.}
\label{fig:1loophard}
  \end{center}
\end{figure}

The leading contribution from non-resonant production-decay
operators ${\cal O}^{(k)}_{4e}$ to (\ref{eq:master}) arises from
four-electron operators of the form
\begin{equation}
\label{eq:4eLag}
{\cal O}^{(k)}_{4e}=
\frac{C^{(k)}_{4e}}{M_W^2}
(\bar e_{c_1}\Gamma_1 e_{c_2})(\bar e_{c_2}\Gamma_2 e_{c_1}),
\end{equation}
where $\Gamma_1$, $\Gamma_2$ are Dirac matrices.
If $C^{(k)}_{4e}\sim \alpha^n$, the contribution to the
forward-scattering amplitude scales as $\alpha^n$. This should
be compared to  ${\cal A}^{(0)}_{LR}\sim \alpha^2\delta^{1/2}$.
The calculation of the short-distance coefficients
$C^{(k)}_{4e}$ is performed in standard fixed-order perturbation
theory in the full electroweak theory. The $W$ propagator is
the free propagator,
since the self-energy insertions are treated perturbatively.
The leading contribution to the forward-scattering amplitude
arises from the one-loop diagrams shown in Figure~\ref{fig:1loophard}.
We will calculate the imaginary part of the short-distance  coefficients
$C^{(k)}_{4e}$ by evaluating the cut diagrams.
The calculation of cuts corresponding to tree amplitudes is
most conveniently performed in unitary gauge with $W$ propagator $-i
(g_{\mu\nu}-k_\mu k_\nu/M_W^2)/(k^2-M_W^2+i \epsilon)$.
To leading order in the expansion in $\delta$, the cut one-loop diagrams
in Figure~\ref{fig:1loophard}
correspond to the production cross section of two on-shell $W$ bosons
directly at threshold, which vanishes. In fact, from an explicit
representation of these one-loop diagrams it can be seen
that the imaginary parts from the hard
region vanish in dimensional regularization to all orders in the
$\delta$ expansion. Thus the leading 
imaginary parts of $C^{(k)}_{4e}$ arise from two-loop
diagrams of order $\alpha^3$. Just as the Coulomb correction
the leading non-resonant (hard) contribution provides another
N$^{1/2}$LO correction relative to~(\ref{eq:Alead}).

\subsection{Classification of corrections up to NLO }
\label{sec:nlo}

We now give an overview of the contributions to the
four-fermion cross section at N$^{1/2}$LO and NLO. These consist of
the short-distance coefficients of the non-relativistic
Lagrangian~\eqref{LPNR}, of the production
operators ${\cal O}^{(k)}_p$, and the four-electron
operators ${\cal O}^{(k)}_{4e}$ on the one hand; and corrections
that arise in calculating the matrix elements in (\ref{eq:master})
within the effective theory on the other.

\subsubsection{Short-distance coefficients in the effective
Lagrangian}
\label{sec:nlo-kinetic}

The effective Lagrangian~\eqref{LPNR} is already complete to
NLO. The only non-trivial matching coefficient is $\Delta$,
which follows from the location of the $W$ pole, which in turn
can be computed from the expansion of the
self-energy~\cite{Beneke:2004km}. In the pole scheme, we
require the NLO correction to the decay width $\Gamma_W$,
defined as the
imaginary part of the pole location, see (\ref{eq:pole}),
(\ref{eq:Deltadef}). At leading order, $\Delta^{(1)} = -i
\Gamma_W^{(0)}$ with\footnote{Here the masses of the light fermions
are neglected, and the CKM matrix has been set to the unit matrix.}
\begin{equation}\label{eq:gamma0}
\Gamma_W^{(0)}=\frac{3}{4} \alpha_{ew} M_W.
\end{equation}

There are electroweak as well as QCD corrections to the $W$
self-energy. We shall count the strong coupling $\alpha_s$ as
$\alpha_{ew}^{1/2}$. Thus the mixed QCD-electroweak two-loop
self-energy provides a N$^{1/2}$LO correction to
$\Delta$, while at NLO we need the self-energy at orders
$\alpha_{ew}^2$ and $\alpha_{ew}\alpha_s^2$.
The QCD effects are included by multiplying the leading-order
hadronic partial decay widths by the universal QCD correction
for massless quarks~\cite{Chetyrkin:1996ia},
\begin{equation}
  \delta_{\rm{QCD}}=1+\frac{\alpha_s}{\pi}+1.409 \,
\frac{\alpha_s^2}{\pi^2},
\label{eq:delta_qcd}
\end{equation}
with $\alpha_s=\alpha_s(M_W)$ in the $\overline{\rm MS}$ scheme.
The electroweak correction to the pole-scheme decay width is
denoted by $\Gamma_W^{(1,ew)}$. The explicit expression is
given in Section~\ref{subsec:hard}. We therefore have
\begin{equation}
\Delta^{(3/2)}=-i \Gamma^{(1/2)}_W  =
-i \,\frac{2\alpha_s}{3\pi}\Gamma_W^{(0)},\qquad
\Delta^{(2)}=-i \Gamma^{(1)}_W  =
-i\left[\Gamma^{(1,ew)}_W+ 1.409
  \,\frac{2\alpha_s^2}{3\pi^2}\Gamma_W^{(0)}\right].
\label{eq:DeltaNLO}
\end{equation}
These results refer to the total width, which appears in the
propagator and the forward-scattering amplitude. The extraction of
the flavour-specific process
$e^- e^+\to\mu^- \bar{\nu}_\mu\, u\, \bar{d}\,X$ will be
discussed in Section~\ref{subsec:NLOwidth}.

\subsubsection{Matching coefficients of the production operators}
\label{sec:match-prod}

There are two sorts of corrections related to production operators:
higher-dimensional operators suppressed by powers of $\delta$, and
one-loop corrections to the operators of lowest dimension such
as (\ref{LPlead}).

The higher-dimension production operators are of the form
\begin{equation}
\label{eq:prod-higher}
{\cal O}_p^{(k)}=\frac{C^{(k)}}{M_W^{2(1+k)}}
  (\bar e_{L/R}\Gamma \mathcal{F}(\vec n,D) e_{L/R})
(\Omega_-^{i\dagger}\mathcal{G}(\vec D)\Omega_+^{j\dagger}),
\end{equation}
where $\Gamma$ is some combination of Dirac matrices and
${\cal F}$ and ${\cal G}$ are functions of the covariant
derivative $D$ acting on the fields. (Here and below, we drop the
collinear direction label on the electron fields, whenever they are
obvious.) The short-distance coefficients
of these operators follow from the expansion of appropriate on-shell
amplitudes around the threshold. The expansion parameter is
$v\sim \delta^{1/2}$. However, for the inclusive cross section
there is no interference of the $v$-suppressed operator with the
leading one, hence the correction from higher-dimension operators
begins at NLO. Full results for the tree-level matching of the
N$^{1/2}$LO production operators are given in~\cite{Beneke:2004xd}.
The NLO contribution to the inclusive cross section is
computed in Section~\ref{subsec:NLOpotential}.

The one-loop correction to the matching coefficient of the production
vertex~\eqref{LPlead} and the related operator with right-handed
electrons requires to calculate the renormalized scattering
amplitudes for $e_L^-e_R^+\to W^+ W^-$ and $e_R^-e_L^+\to W^+ W^-$ to NLO in
ordinary weak coupling perturbation theory for the momentum configuration
$(p_1+p_2)^2=4M_W^2$, i.e.~directly at threshold.
This generates the NLO production operator
\begin{equation}
  {\cal O}_p^{(1)} = \frac{\pi\alpha_{ew}}{M_W^2} \left[C_{p,LR}^{(1)}
    \left(\bar{e}_L \gamma^{[i} n^{j]} e_L \right) +
    C_{p,RL}^{(1)}
  \left(\bar{e}_R \gamma^{[i} n^{j]} e_R \right)\right]
    \left(\Omega_-^{\dagger i} \Omega_+^{\dagger j}\right) .
    \label{LPNLO}
\end{equation}
The calculation of the coefficients $C_{p,LR}^{(1)}$,
$C_{p,RL}^{(1)}$ is discussed in
Section~\ref{subsec:hard}. Note, however, that
the one-loop correction $C_{p,RL}^{(1)}$ does in fact not
contribute to the NLO cross section, since there is
no leading-order contribution from the  $e_R^-e_L^+$ helicity
initial state, and no interference between LR and RL
configurations.

\subsubsection{Matching coefficients of four-electron operators}
\label{sec:match-hard-onehalf}

As discussed above the leading contributions from the non-resonant
production-decay operators to the imaginary part of the forward scattering
amplitude arise at N$^{1/2}$LO, where the
half-integer scaling arises from the absence of the threshold
suppression $v\sim \delta^{1/2}$ present in the LO cross section.
The calculation of the cut 2-loop diagrams amounts to the
calculation of the squared and phase-space integrated
matrix element of the on-shell processes $e^-e^+\to W^-u\bar d$
and $e^-e^+\to \mu^-\bar \nu_\mu W^+ $
in ordinary perturbation theory (no ``resummations'' in
internal $W$ propagators). This includes contributions
of what is usually called double-resonant (or CC03)
diagrams, where one of the $W$ propagators is in fact off-shell,
as well as genuine single-resonant processes. In the terminology of
the method of regions, these corrections are
given by the hard-hard part of the two-loop forward-scattering amplitude.
Since they contain all diagrams contributing to the tree-level
scattering processes $e^-e^+\to \mu^-\bar\nu_\mu W^+$ and $e^-
e^+\to W^- u\bar d$, the matching coefficients are gauge invariant.
Since only one $W$ line is cut in the
N$^{1/2}$LO contributions, they can be viewed as systematic
corrections to the narrow-width approximation. This calculation
is performed in Section~\ref{subsec:N12LOhard}.

To NLO in the power counting $\alpha_s^2\sim\alpha_{ew}$ we would have
to compute also the NLO QCD corrections to $e^- e^+ \to W^- u \bar d \,(+g)$.
The corrections to the ``double-resonant'' (CC03) diagrams
can be taken into account approximately by multiplying them
with the one-loop QCD correction to the hadronic decay width.
The corrections to the single-resonant diagrams require the full
calculation. However, we shall find that the contribution
of the single-resonant diagrams to $e^-e^+\to W^-u\bar d$ is
numerically already small, so we neglect the QCD corrections.

\subsubsection{Calculations in the effective theory}

\paragraph{\it One-loop diagrams with insertions of subleading
  operators.} The contributions in this class arise from evaluating
the first term in~\eqref{eq:master} at one loop, see
Figure~\ref{fig:LO}, but with one insertion of the subleading
bilinear terms in the Lagrangian (\ref{LPNR}), which correspond to
kinetic energy and width corrections, or with production
operator products ${\cal O}_p^{(0)}{\cal O}_p^{(1)}$ and ${\cal
 O}_p^{(1/2)}{\cal O}_p^{(1/2)}$, where ${\cal O}_p^{(1)}$
is either a higher-dimension operator (\ref{eq:prod-higher})
or the one-loop correction (\ref{LPNLO}).
As already mentioned the N$^{1/2}$LO products
${\cal O}_p^{(0)}{\cal O}_p^{(1/2)}$ vanish after performing the
angular integrals. In the calculation
discussed further in Section~\ref{sec:tree} we actually follow a
different approach and directly expand the spin-averaged squared
matrix elements rather than the amplitude before squaring, which
would yield the individual production vertices.

\paragraph{\it Coulomb corrections.} A single insertion
of the Coulomb potential interaction in the
Lagrangian (\ref{LPNR}) contributes at N$^{1/2}$LO. To NLO one
has to calculate the double insertion into the leading-order
amplitude from ${\cal O}_p^{(0)}{\cal O}_p^{(0)}$ and a single
insertion into ${\cal O}_p^{(0)}{\cal O}_p^{(1/2)}$. The latter
vanishes for the total cross section. There is no coupling of the
potential photons to the collinear electrons and positrons, so
there are no Coulomb corrections to the four-fermion operators.
The Coulomb corrections are given in Section~\ref{subsec:coulomb}.

\paragraph{\it NLO corrections from soft and collinear photons.}
To NLO one has to calculate two-loop diagrams in the effective theory
arising from the coupling of the collinear modes and the potential $W$
bosons to the soft and collinear photons contained in the NRQED
Lagrangian~\eqref{LNR} and the SCET Lagrangian.  The cuts correspond
to one-loop virtual and bremsstrahlung corrections to the leading-order
cross section. In the terminology of the method of regions these are
contributions from the soft-potential, the $c_1$-potential and the
$c_2$-potential regions. They
correspond to ``non-factorizable corrections'' and are discussed in
Section~\ref{subsec:soft}.

\section{Expansion of the Born cross section}
\label{sec:tree}

This section serves two purposes. First, we calculate
all NLO corrections to four-fermion production in the
effective theory (EFT) except those related to loop corrections,
which will be added in Section~\ref{sec:radcor}. Second, we
investigate the convergence of the successive EFT approximations
to what is usually referred to as the Born four-fermion
production cross section. The two calculations are not exactly
the same, since the implementation of the $W$ width in
the Born cross section is not unique. We {\em define} the
``exact'' Born cross section by the ten tree diagrams for $e^{-} e^{+}
\rightarrow \mu^{-} \bar{\nu}_\mu u \bar{d}$, where the $W$ propagators
are supplied with a fixed-width prescription. The EFT calculation is
done by expanding directly the forward-scattering amplitude.
The relevant loop momentum regions are either all hard, or
hard and potential. In the latter regions the two $W$ propagators
and the $W$ interactions are described by the non-relativistic
Lagrangian. The all-hard contributions correspond to the
matching and matrix element of the four-electron operators.

\subsection{Expansion in the potential region}
\label{subsec:NLOpotential}

We first reconsider the one-loop diagrams (before cutting) shown in
Figure~\ref{fig:1loophard},  where the loop momentum  is
now assumed to be in the potential region. The forward-scattering amplitude
corresponding to these diagrams may be written as
\begin{equation}
\label{eq:amppot}
i \mathcal{A} = \int \frac{d^d r}{(2 \pi)^d} \,\Phi(E, r)
P(k_1) P(k_2),
\end{equation}
where $E=\sqrt{s}-2 M_W$, $k_1=M_W v+r$, $k_2=P-M_W v-r$,
with $v^{\mu}=(1,\vec{0}\,)$
and $P=p_1+p_2$ the sum of the initial-state momenta. Here
$\Phi(E,r)$ is the square of the off-shell
$W$ pair-production amplitude at tree level, including the
numerator $(-g_{\mu\nu}+k_\mu k_\nu/k^2)$ from the
$W$ propagators, and
\begin{equation}
\label{eq:Wresum}
P(k)=\frac{i}{k^2-M_W^2-\Pi_T^W(k^2) }
\end{equation}
is the full renormalized (transverse) $W$
propagator.\footnote{The longitudinal
part of the propagator is cancelled by the transverse projector
from the decay into massless fermions.} Writing the amplitude
in the full theory with a resummed propagator is contrary to the
spirit of effective field theory calculations, where the matching
coefficients are obtained by fixed-order calculations. However,
this allows us to compare the EFT expansion
with the standard calculation of the fixed-width Born cross section.

To see the correspondence with the EFT calculation, we parameterize
the $W$ momentum as $k^{\mu} = M_W v^{\mu}+r^{\mu}$, where
$r^{\mu}$ is a potential residual
momentum ($r_0 \sim M_W \delta$, $\vec{r} \sim M_W\delta^{1/2}$),
and expand $P(k)$ in $\delta$, including an expansion of
the self-energy around $M_W^2$ and in the number of loops,
\be
\Pi_T^W(k^2) = M_W^2 \sum_{m,n}  \delta^n \, \Pi^{(m,n)},
\label{eq:Pihard}
\ee
with $\delta=(k^2-M_W^2)/M_W^2$ and $m$ denoting the loop order.
The result is
\begin{equation} \label{eq:Wpotprop}
P(r) = \frac{i(1+\Pi^{(1,1)})}{2 M_W \left(r_0-\frac{\vec{r}^{\,2}}{2
M_W}- \frac{\Delta^{[1]}}{2}\right)}
-\frac{i(r_0^2- M_W\Delta^{(2)})}{4
M_W^2\left(r_0-\frac{\vec{r}^{\,2}}{2 M_W}-
\frac{\Delta^{[1]}}{2}\right)^2}+O\!\left(\frac{\delta}{M_W^2}\right),
\end{equation}
where, to make the notation simpler, we included the QCD
correction $\Delta^{(3/2)}$ from (\ref{eq:DeltaNLO}) into
$\Delta^{[1]}=\Delta^{(1)}+\Delta^{(3/2)}$ instead of expanding it
out, and $\Delta^{(2)}=M_W(\Pi^{(2,0)}+\Pi^{(1,1)}\Pi^{(1,0)})$.
Next we eliminate $r_0$ from the numerator
in~\eqref{eq:Wpotprop} by completing the square and obtain
\begin{eqnarray}
P(r)&=& \frac{i}{2 M_W \left(r_0-\frac{\vec{r}^{\,2}}{2M_W}
- \frac{\Delta^{[1]}}{2}\right)}
\left(1+\Pi^{(1,1)}-\frac{M_W\Delta^{[1]}+\vec r^{\,2}
  }{2M^2_W}\right)
\nonumber\\
&& -\,
\frac{i\left[\left(\frac{\vec r^{\,2}}{2M_W} +\frac{\Delta^{[1]}}{2}\right)^2
 - M_W\Delta^{(2)}\right]}{4M_W^2\left(r_0-\frac{\vec{r}^{\,2}}{2 M_W}-
\frac{\Delta^{[1]}}{2}\right)^2}-\frac{i}{4M_W^2}
+O\!\left(\frac{\delta}{M_W^2}\right).
\label{eq:NLO-prop}
\end{eqnarray}
The individual terms now have a clear interpretation in the
EFT formalism. The first term in the second line corresponds to
a single insertion of the NLO terms -- a kinetic energy correction
and a second-order width correction -- in the non-relativistic
Lagrangian (\ref{LPNR}) into a $W$ line. The local term,
$-i/(4 M_W^2)$, in the second line is similar to a corresponding
term in single resonance production~\cite{Beneke:2004km}, where it
contributes to a production-decay vertex at tree level. Here this
term leads to potential loop integrals with only one or no
non-relativistic $W$ propagator, which vanish in dimensional
regularization. Thus, we can drop this term. In the first line
of (\ref{eq:NLO-prop}) we recognize the non-relativistic
$W$ propagator (\ref{OmegaProp}) multiplied by a correction to
the residue. The residue correction originates from the expansion
of the field normalization factor $\varpi$ defined
in (\ref{eq:Phinorm}), and from the derivative of the renormalized
one-loop self-energy, $\Pi^{(1,1)}$, at $k^2=M_W^2$. In an EFT
calculation these residue corrections are not associated with
the propagator, but they enter the matching relations of the
one-loop and higher-dimension production and decay
vertices~\cite{Beneke:2004km}. In order to compare
with the ``exact'' Born cross section, where these terms are included,
we keep these residue corrections here rather than in the matching
calculation of Section~\ref{subsec:hard}.

The real part of $\Pi^{(1,1)}$
depends on the $W$ field-renormalization convention in the
full theory. In the following we adopt the on-shell scheme for
field renormalization, $\mbox{Re}\,\Pi^{(1,1)}=0$, and the pole
scheme for mass renormalization. Since $\mbox{Im} \,\Pi_T^W(k^2) =
-k^2 \Gamma_W^{(0)}/M_W \,\theta(k^2)$ at one-loop due to the decay into
massless fermions, it follows
that $\Pi^{(1,1)} = -i\Gamma_W^{(0)}/M_W$. Furthermore,
$\Delta^{(1)}=M_W \Pi^{(1,0)} = -i \Gamma_W^{(0)}$ and
$\Delta^{(2)}=M_W (\Pi^{(2,0)}+\Pi^{(1,1)}\Pi^{(1,0)}) =
-i \Gamma_W^{(1)}$ in the pole mass renormalization scheme,
which implies $\mbox{Re}\,\Pi^{(2,0)} = (\Gamma_W^{(0)}/M_W)^2$,
$\mbox{Im}\,\Pi^{(2,0)} = -\Gamma_W^{(1)}/M_W$ for the
renormalized two-loop self-energy at $k^2=M_W^2$. The QCD
correction $\Delta^{(3/2)}=-i \Gamma_W^{(1/2)}$ can
be included into $-i \Gamma_W^{(0)}$ as before.

To compare with the ``exact'' Born cross section, we write
(\ref{eq:Wresum}) in this renormalization scheme in the form
\be
P(k) = i\,\frac{k^2-M_W^2-{\Gamma_W^{(0)}}^2-i M_W
\left(k^2 \Gamma_W^{(0)}/M_W^2+\Gamma_W^{(1)}\right)}
{\left(k^2-M_W^2-{\Gamma_W^{(0)}}^2\,\right)^2+M_W^2
\left(k^2 \Gamma_W^{(0)}/M_W^2+\Gamma_W^{(1)}\right)^2}
+O\left(\frac{\delta}{M_W^2}\right).
\ee
The fixed-width prescription corresponds to replacing
$k^2 \Gamma_W^{(0)}/M_W^2$ by $\Gamma_W^{(0)}$ in
the denominator, but not in the numerator, where the
factor of $k^2$ arises from the integration over the
two-particle phase space of the $W$ decay products.
In addition one drops the ${\Gamma_W^{(0)}}^2$ terms 
(since they come from $\mbox{Re}\,\Pi^{(2,0)}$) and
$\Gamma_W^{(1)}$. Repeating the derivation
of (\ref{eq:NLO-prop}) with this modified expression
we obtain
\be
P(k)_{\rm fixed-width} =
\Big[\mbox{Eq. (\ref{eq:NLO-prop}) with $\Delta^{(2)}=-i \Gamma_W^{(1)}
\to 0$}\Big]
+ \frac{{\Gamma_W^{(0)}}^2}
{\left(k^2-M_W^2\right)^2+M_W^2 {\Gamma_W^{(0)}}^2} \,.
\ee
The additional term is purely real and does not contribute
to the cut propagator $\mbox{Im}\,P(k)$ relevant to the
cross-section calculation. We therefore arrive at the interesting
conclusion that the fixed-width prescription coincides
with the EFT approximation in the potential region up to the
next-to-leading order, if $M_W$ is the pole mass, up to
a trivial term related to the one-loop correction $\Gamma_W^{(1)}$
to the pole scheme decay width.

In the calculation of the NLO correction to the
forward-scattering amplitude in the potential region, we
use the expansion \eqref{eq:Wpotprop} in (\ref{eq:amppot}),
and drop all terms beyond NLO. This already accounts for
all NLO corrections from the effective Lagrangian, and
for some corrections from higher-dimension production operators
with tree-level short-distance coefficients. Further corrections
of this type come from the expansion of the squared
matrix element $\Phi(E,r)$. The square of the production
amplitude of two off-shell $W$ bosons depends on four
kinematic invariants, which we may choose to be $r^2$,
$p_1\cdot r$, $k_1^2-M_W^2$, and $k_2^2-M_W^2$. This choice
is convenient, since all four invariants are small with respect
to $M_W^2$ in the potential region. In the expansion of
$\Phi(E,r)$ to NLO, we may further approximate $r^2$ by
$-\vec{r}^{\,2}$, since  $r_0\sim \vec{r}^{\,2}/M_W\ll
|\vec{r}\,|$ and exploit that $P(k_{1,2})$ does not depend
on the direction of $\vec{r}$. We find, for the  $e^-_Le^+_R$
and  $e^-_Re^+_L$ helicity initial states (the LL and RR
combinations vanish),
\begin{eqnarray}
\Phi_{LR}(E,r)&=&- 64 \pi^2 \alpha_{ew}^2 \left[1+\left(\frac{11}{6}+2
\xi^2(s)+\frac{38}{9}
\xi(s)\right)\frac{\vec{r}^{\,2}}{M_W^2}\right] +O(\delta^2),
\nonumber\\
\Phi_{RL}(E,r)&=&- 128 \pi^2 \alpha_{ew}^2\,
\chi^2(s)\frac{\vec{r}^{\,2}}{M_W^2} +
O(\delta^2).
\label{eq:matrixpot}
\end{eqnarray}
The functions
\be
\xi(s) = - \frac{3 M_W^2 (s-2 M_Z^2 s_w^2)}{s (s-M_Z^2)},
\qquad
\chi(s) = -\frac{6 M_W^2 M_Z^2 s_w^2}{s (s-M_Z^2)}
\label{xichi}
\ee
originate from the $s$-channel photon and $Z$ boson propagators.
The NLO terms proportional to $\vec{r}^{\,2}$ can be identified
with tree-level production
operator products ${\cal O}_p^{(0)}{\cal O}_p^{(1)}$ and ${\cal
 O}_p^{(1/2)}{\cal O}_p^{(1/2)}$ as discussed in
Section~\ref{sec:nlo}. In such calculations $\xi(s)$ and
$\chi(s)$ would be evaluated at $s=4M_W^2$. Here we keep
the exact $s$-dependence, since this can be done at no
calculational cost.

Note that the coefficient functions of production operators in
the EFT are determined by on-shell matching, which implies
an expansion of amplitudes around the complex pole position
$\bar s=M_W^2+M_W \Delta$ rather than
$M_W^2$~\cite{Aeppli:1993rs,Stuart:1991xk}. The difference
cannot be neglected in NLO calculations. In principle the
expansions (\ref{eq:matrixpot}) could have yielded terms
such as $k_1^2-M_W^2$, which should be written as
$k_1^2-\bar s +M_W \Delta$. The difference $k_1^2-\bar s$
cancels a resonant propagator (possibly giving rise to a
production-decay operator matching coefficient), while the
remaining $M_W \Delta$ term must be combined with other contributions to
the loop correction to the leading-order production vertex.
This complication can be ignored here, since the expansion
of $\Phi(E,r)$ is independent of $k_{1,2}^2-M_W^2$ up to NLO.

The NLO correction from the potential region is now obtained
by inserting the expansions
(\ref{eq:Wpotprop}), (\ref{eq:matrixpot}) into
(\ref{eq:amppot}) and performing the loop integral.
The integral has an odd power-divergence which is
finite in dimensional regularization. The LO
cross section has already been given in~\eqref{eftLOsigma}.
The NLO terms are
\begin{eqnarray}
\sigma^{(1)}_{LR,\mbox{\tiny Born}} &=& \frac{4\pi \alpha^2}{27 s_w^4 s} \,
\Bigg\{\left( \frac{11}{6}+2\xi^2(s)+\frac{38}{9} \xi(s)\right)
\mbox{Im}
\Bigg[\left(-\frac{E+i \Gamma_W^{(0)}}{M_W}\right)^{\!3/2}\Bigg]
\nonumber\\
&&\hspace*{1.2cm}+\,\mbox{Im} \Bigg[
\left(\frac{3 E}{8 M_W}
+\frac{17\, i \Gamma_W^{(0)}}{8M_W}\right)
\sqrt{-\frac{E+i \Gamma_W^{(0)}}{M_W}}\nonumber\\
&&\hspace{2.3 cm}
-\left(\frac{{ \Gamma_W^{(0)}}^2}{8 M_W^2}-
\frac{i \Gamma_W^{(1)}}{2 M_W}\right)
\sqrt{-\frac{M_W}{E +i \Gamma_W^{(0)}}} \,\Bigg]
\Bigg\},
\nonumber\\
\sigma^{(1)}_{RL,\mbox{\tiny Born}}&=&
\frac{8\pi \alpha^2}{27 s_w^4 s}\,\chi^2(s)\,
\mbox{Im}\Bigg[\left(-\frac{E+i \Gamma_W^{(0)}}{M_W}\right)^{\!3/2}\Bigg] .
\label{eftNLOpotsigma}
\end{eqnarray}
Since $E/M_W\sim \Gamma_W^{(0)}/M_W\sim \delta$ and
$\Gamma_W^{(1)}/M_W\sim \delta^2$ every term is suppressed by
$\delta$ relative to the leading order as it should be. The
unpolarized cross section is one fourth the sum of the LR, RL
contributions. The factor 1/27 comes from the tree-level branching
ratio for the final state  $\mu^{-} \bar{\nu}_\mu \,u \bar{d}$ in
the conversion from the forward-scattering amplitude to the
partial cross section. As discussed above, when we use this
expression to compare with the standard Born cross section in the
fixed-width scheme, we set $\Gamma_W^{(1)}$ to zero. When we use
the expression (\ref{eftNLOpotsigma}) in the complete NLO
calculation including radiative corrections, we have to keep in
mind that multiplying all terms by the product $1/27$ of
leading-order branching fractions as in (\ref{eftNLOpotsigma}) is
actually not correct. The required modification is discussed in
Section \ref{subsec:NLOwidth}.

\begin{figure}[t]
\begin{center}
\includegraphics[width=0.35 \linewidth]{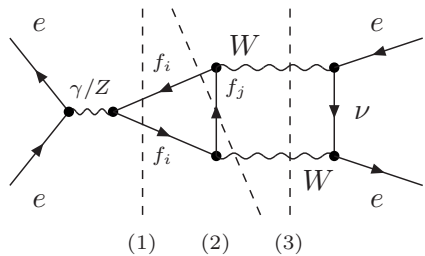}
\end{center}
\caption{Example of a two-loop diagram with one hard and one potential
loop. Cut (2) is part of the Born cross section, but subleading as
discussed in the text.}
\label{fig:potcuts}
\end{figure}

In addition to the $\delta$-suppressed terms from the potential
region of the one-loop diagrams shown in Figure~\ref{fig:1loophard},
there is another NLO contribution from the leading terms of
two-loop diagrams with one hard and one potential loop, which
may also be associated with the Born cross section. An example
is displayed in Figure~\ref{fig:potcuts}. Cut (1) does not
correspond to a four-fermion final state and must be dropped.
Cut (3) corresponds to the interference of a tree-level production
operator with the real part of
a hard one-loop correction to a production operator.
Since the $s$-channel diagrams do not contribute to the leading-power
production operator, this cut is beyond NLO. Cut (2) is a
contribution to what is usually termed the ``Born cross section''
corresponding to the interference of single and double resonant
diagrams in the kinematic region where both fermion pairs
have invariant mass of order $M_W^2$. The contribution from this
cut is contained in the imaginary parts of the
hard one-loop correction to the production operators.
The threshold suppression
of the $s$-channel diagrams applies here as well, hence this
contribution is also not relevant at NLO.

\subsection{Decay-width correction for the flavour-specific
cross section}
\label{subsec:NLOwidth}

As already noted, the expression (\ref{eftNLOpotsigma}) has to be modified
in order to take the radiative correction to the decay correctly into
account. In this subsection we derive the
required modification of the formula, but note that it will
not be needed for the comparison to the Born cross section,
where radiative corrections are excluded.

To include the loop corrections to $W$ decay
for the flavour-specific four-fermion
final state $\mu^{-} \bar{\nu}_\mu \,u \bar{d}$ we have to identify
contributions to the forward-scattering amplitude from cut two-loop
$W$ self-energy insertions and include only the appropriate cuts
containing a muon and muon-antineutrino or up and anti-down quarks and,
possibly, a photon.
Repeating the expansion in the potential region
performed in Section~\ref{subsec:NLOpotential}
for the cut diagram with
flavour-specific cuts selected, one finds that in the pole
mass renormalization and on-shell field renormalization scheme
adopted here
all terms in the expansion are correctly treated by multiplying the
totally inclusive
result by the ratio of leading-order partial branching fractions,
$\Gamma^{(0)}_{\mu^-\bar\nu_\mu}\Gamma_{u\bar d}^{(0)}/
[\Gamma^{(0)}_W]^2=1/27$, except for one term involving the
insertion of $\Delta^{(2)}=-i
\Gamma_W^{(1)}$. In~\eqref{eftNLOpotsigma} this insertion
results in
part of the term involving $\Gamma^{(1)}_W$, and is also multiplied by
$1/27$. We therefore have to modify this
term to include the flavour-specific cuts correctly.
At NLO we have to consider diagrams where $ i\Delta^{(2)}/2$ is inserted
in only one of the two $W$-lines. Cutting this line produces
a contribution to the imaginary part of the forward-scattering
amplitude of the form
\begin{equation}
\label{eq:cut-line}
\im\left[(-i) \frac{i}{\eta}\frac{i\Delta^{(2)}}{2} \frac{i}{\eta}\right]
=- \im\left[\frac{1}{\eta}\right]
\left(\frac{\Delta^{(2)}}{2}\right)^{\!\!*} \frac{1}{\eta^*}
- \frac{1}{\eta}
\frac{\Delta^{(2)}}{2}\im \left[\frac{1}{\eta}\right]
- \frac{1}{\eta} \left[ \frac{\im \Delta^{(2)}}{2}\right]
\frac{1}{\eta^*}
\end{equation}
where $\eta$ is the inverse propagator of the non-relativistic $W$ boson.
The first two terms correspond to cutting the $W$ line to the left and right
of the $\Delta^{(2)}$ insertion. The flavour-specific final states
are extracted from these  cuts as discussed
below~\eqref{eq:im-propagator}. This amounts to multiplying the NLO
correction~\eqref{eftNLOpotsigma} by the leading-order branching
ratios, so these two terms are treated correctly by the factor $1/27$.
The last term corresponds to a cut two loop self-energy insertion,
where only the cuts leading to the desired final state must be taken
into account.  Therefore here $-\im \Delta^{(2)}=\Gamma_W^{(1)}$
has to be replaced by
$\Gamma^{(1)}_{\mu^-\bar\nu_\mu}=
\Gamma^{(1,ew)}_{\mu^-\bar\nu_\mu}$
and $\Gamma^{(1)}_{u\bar d} =\Gamma^{(1,ew)}_{u\bar d}+  1.409
  \,\frac{\alpha_s^2}{\pi^2}\, \Gamma_{u\bar d}^{(0)}$,
respectively, to obtain
the NLO cross section for the four-fermion final state.
To implement these replacements, note that
the contribution of the last term
in~\eqref{eq:cut-line} to the forward-scattering amplitude is
of the form $\Gamma^{(1)}_W/\Gamma^{(0)}_W \,\im \,{\cal A}^{(0)}$.
We can therefore compensate the incorrect treatment of
the flavour-specific cross section in~\eqref{eftNLOpotsigma} by
subtracting this contribution for each
$W$ line and adding the flavour-specific corrections.
Multiplying by the leading-order branching
fraction for the second $W$ line one obtains
the additional NLO correction to the cross section,
\begin{align}
  \Delta\sigma^{(1)}_{\text{decay}}&=
\left(\frac{\Gamma^{(1)}_{\mu^-\bar\nu_\mu}}{{
\Gamma^{(0)}_{\mu^-\bar\nu_\mu} }}+
\frac{\Gamma_{u\bar d}^{(1)}}{\Gamma_{u\bar d}^{(0)} }
-2\,\frac{\Gamma_W^{(1)}}{\Gamma_W^{(0)}}  \right) \sigma^{(0)} .
\label{eq:decaycorr}
\end{align}
At NLO this correction is equivalent to
multiplying the imaginary part of the
leading-order (or even next-to-leading order) forward-scattering
amplitude by the one-loop corrected branching ratios
$\Gamma^{(\rm NLO)}_{\mu^-\bar\nu_\mu} \Gamma_{u\bar
  d}^{(\rm NLO)}/ [\Gamma^{(\rm NLO)}_W]^2$ rather than
by $1/27$, where $\Gamma^{(\rm NLO)}_X=\Gamma^{(0)}_X+\Gamma_X^{(1)}$.
The NLO partial decay rates are
calculated in Section~\ref{subsec:hard}.

\subsection{Expansion in the hard region}
\label{subsec:N12LOhard}

\begin{figure}[t]
\begin{center}
\includegraphics[width=0.7 \linewidth]{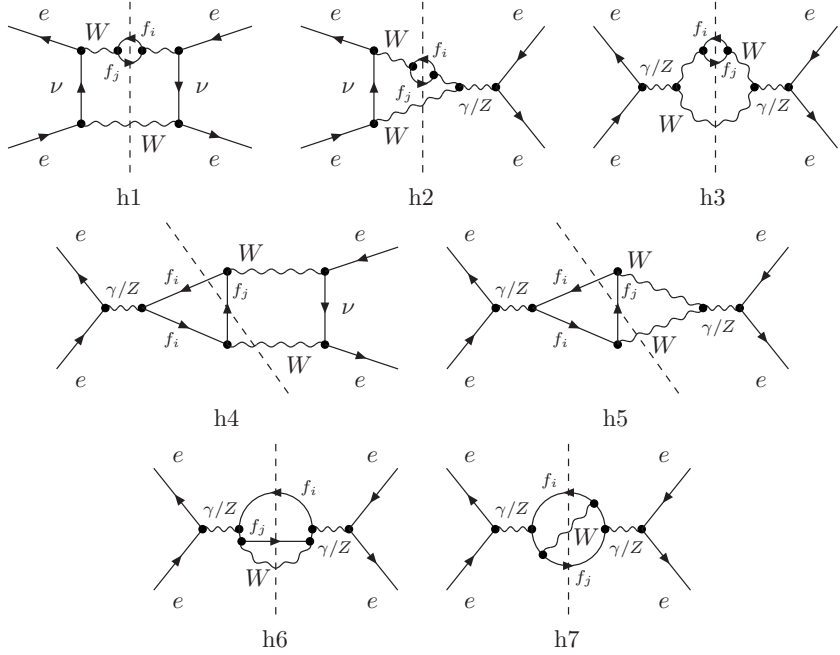}
\end{center}
\caption{Two-loop cut diagrams. Symmetric diagrams are not shown.}
\label{fig:hardcuts}
\end{figure}

We now consider the hard contributions, which determine the
matching coefficients of four-electron production-decay operators.
As already discussed in Section~\ref{sec:production}, the one-loop
diagrams shown in Figure~\ref{fig:1loophard} do not provide
imaginary parts of the forward-scattering amplitude. The leading
hard contributions originate from the two-loop diagrams in
Figure~\ref{fig:hardcuts}. These diagrams are to be calculated in
standard perturbation theory with no width added to the $W$
propagator, but expanded near threshold. The result must be of
order $\alpha^3$, which results in a N$^{1/2}$LO correction
relative to the leading-order cross section. Higher-order terms in
the hard region come from higher-order terms in the expansion (in
$E=\sqrt{s}-2 M_W$) near threshold and from diagrams with more
hard loops, all of which are N$^{3/2}$LO and smaller.

In the hard region it is simpler to calculate the four-fermion
cross section directly as the sum over the relevant cuts of the
forward-scattering amplitude as shown in Figure~\ref{fig:hardcuts}.
Note that this includes cutting $W$ lines as well as diagrams
with self-energy insertions into the $W$ propagator. This
can be interpreted as an expansion of the resummed propagator
in the distribution sense \cite{Tkachov:1998uy,Nekrasov:2002mw}, such as
\begin{equation}
\label{eq:prophard}
\frac{M_W\Gamma_W }{(k^2-M_W^2)^2+ M_W^2\Gamma_W^2}
=\pi \delta(k^2-M_W^2)+\mbox{PV}\,
\frac{M_W\Gamma_W}{(k^2-M_W^2)^2}+O\left(\frac{\delta^2}{M_W^2}\right),
\end{equation}
``PV'' denoting the principal value. The left-hand side arises
from cutting fermion-loop insertions into the $W$ propagator, but not
the $W$ lines itself. But the leading term in the expansion
of this expression, equivalent to the narrow-width approximation,
looks as if a $W$ line with no self-energy insertions is cut.

The principal-value prescription is redundant at N$^{1/2}$LO, where
the singularity in the integrand is
located at one of the integration limits, and is regularized by
dimensional regularization, which has to be supplied in any case
to regulate infrared divergences that arise as a consequence of
factorizing hard and potential regions in the threshold
expansion. As in the potential region, the integrals are
actually analytically continued to finite values, since the
divergences are odd power divergences.
The result of the calculation can be written as
\begin{align}
\sigma^{(1/2)}_{LR,\mbox{\tiny Born}}&=\frac{4 \alpha^3}{27 s_w^6 s}
\left[ K_{h1} +K_{h2}\, \xi(s) + K_{h3}\, \xi^2(s)
+\sum_{i=h4}^{h7} \sum_f
C^{f}_{i,LR}(s)
K^{f}_i\right],\nonumber \\
\sigma^{(1/2)}_{RL,\mbox{\tiny Born}}&=\frac{4 \alpha^3}{27 s_w^6 s}
\left[  K_{h3}\,\chi^2(s)
+\sum_{i=h4}^{h7} \sum_f
C^{f}_{i,RL}(s)
K^{f}_i\right].
\label{eq:treehard}
\end{align}
Here the first sum extends over the
diagrams as labelled in Figure~\ref{fig:hardcuts}, the
second over the fermions $f\in {u,d,\mu,\nu_\mu}$ in the internal
fermion loops.
The explicit values of the coefficients arising from the diagrams h1-h3
are
\begin{equation}
K_{h1}= -2.35493\, ,\quad
K_{h2}= 3.86286\, ,\quad
K_{h3}= 1.88122.
\end{equation}
The three coefficients contain the contribution of the diagrams
h1-h3 shown in
Figure~\ref{fig:hardcuts} and of the symmetric diagrams with self-energy
insertions on the lower $W$ line. $K_{h2}$ contains also the
contribution of the complex conjugate of h2.
The explicit expressions of coefficients
$K_i^f$ and $C_{i,h}^f$,  with  $h=LR,RL$, for the diagrams h4-h7
are given in Appendix~\ref{ap:4ferm}.
Similar to (\ref{xichi}) the $s$-dependence of the
$C_{i,h}^f$ arises trivially from photon and $Z$ propagators,
and we could put $s=4 M_W^2$ at N$^{1/2}$LO. Since all other
terms in (\ref{eq:treehard}) are energy-independent, we conclude that
the leading hard contribution results in a constant N$^{1/2}$LO shift
of the cross section.

This contribution can be interpreted as arising from a final
state where one fermion pair originates from a nearly on-shell
$W$ decay, while the other is produced non-resonant\-ly, either
from a highly virtual $W$, or as in the truly single-resonant
diagrams h4-h7. Numerical investigation reveals that
the contribution from h4-h7 is rather small, below $0.5\%$ of the
full tree cross section in the energy range $\sqrt s=155\,$GeV
and $180\,$GeV. Below $155\,$GeV it becomes negative and its
magnitude grows to $4\%$ at $150\,$GeV.  The smallness of the single-resonant
contributions is in part due to large cancellations between the
diagrams h4 and h5.

The comparison with the Born cross section performed below shows
that the region of validity of the EFT expansion is significantly
enlarged, if the energy-dependent  N$^{3/2}$LO terms are included.
These can only arise from the next-to-leading order terms of the
expansion in the hard region (the expansion in the potential
region produces only integer-power corrections in $\delta$). The
energy-dependent terms are related to the next order in the
threshold expansion of the cut diagrams in
Figure~\ref{fig:hardcuts}. The computation for the numerically
dominant diagrams h1-h3 gives
\begin{align}
\sigma^{(3/2),a}_{LR,\mbox{\tiny Born}}&=\frac{4 \alpha^3 E}{27 s_w^6 s M_W}
\left[ K_{h1}^a +K_{h2}^a\, \xi(s) + K_{h3}^a\, \xi^2(s) \right],
\nonumber\\
\sigma^{(3/2),a}_{RL,\mbox{\tiny Born}}&=
\frac{4 \alpha^3 E}{27 s_w^6 s M_W} K_{h3}^a\,
\chi^2(s),
\label{eq:treehardNLO}
\end{align}
where
\begin{equation}
K_{h1}^a= -5.87912\, ,\quad
K_{h2}^a= -19.15095\, ,\quad
K_{h3}^a= -6.18662.
\end{equation}
Other N$^{3/2}$LO corrections related to the Born cross section
arise from cut three-loop diagrams of the type h1-h3, but with
two self-energy insertions, and of type h4-h7 with one insertion.
This N$^{3/2}$LO term is (almost) energy-independent and
can be parameterized by
\begin{equation}
\label{eq:n3/2_b}
\sigma^{(3/2),b}_{h,\mbox{\tiny Born}}=
\frac{4 \alpha^4}{27 s_w^8 s} \sum_{i=h1}^{h3}
C_{i,h}^b(s) K_i^b.
\end{equation}
The coefficients $C_{i,h}^b(s)$ are equal to the factors
multiplying $K^a_{hi}$ in (\ref{eq:treehardNLO}) and we omitted the
small contributions from h4-h7. The calculation of
the numerical coefficients $K_i^b$ is non-trivial, since it contains
products of distributions. A rough estimate of these corrections
is $\sigma^{(3/2),b}_{h,\mbox{\tiny Born}} \sim
\sigma^{(1/2)}_{h,\mbox{\tiny Born}}\,
\Gamma_W^{(0)}/M_W \sim 0.025 \,\sigma^{(1/2)}_h$, resulting
in an energy-independent contribution to the cross section of
order $2\,\mbox{fb}$. The comparison below suggests that actually
it is significantly smaller.

\subsection{Comparison to the four-fermion
Born cross section}
\label{subsec:compareBorn}

\begin{figure}[t]
\begin{center}
\includegraphics[width=0.75 \linewidth]{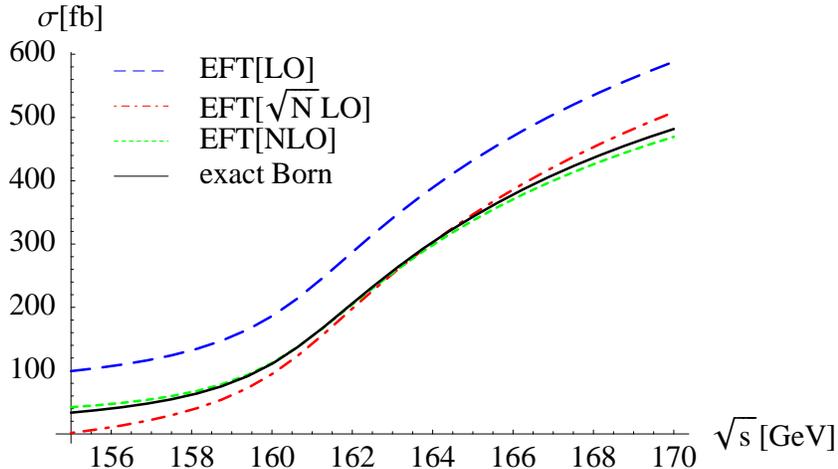}
\end{center}
\vspace*{-0.4cm}
\caption{Successive EFT approximations:
LO (long-dashed/blue), $\mbox{N}^{1/2}$LO (dash-dotted/red) and 
NLO (short-dashed/green). The solid/black 
curve is the full Born result computed with Whizard/CompHep.
The $\mbox{N}^{3/2}$LO EFT approximation is indistinguishable from
the full Born result on the scale of this plot. }
\label{fig:eft}
\end{figure}

We compare the successive EFT approximations to the four-fermion
Born cross section in the fixed-width scheme. We discuss only the
unpolarized cross section given by $(\sigma_{LR}+\sigma_{RL})/4$.
The relevant terms are given in (\ref{eftLOsigma}),
(\ref{eftNLOpotsigma}), (\ref{eq:treehard}), and
(\ref{eq:treehardNLO}).
The input parameters are taken to be
\begin{equation}
\begin{aligned}
\hat{M}_W&=80.403\,\text{GeV} ,&
M_Z&= 91.188\,\text{GeV},&
G_\mu&=1.16637\cdot 10^{-5}\,\text{GeV}^{-2}.
\label{inputs}
\end{aligned}
\end{equation}
The pole mass $M_W$ is related to the on-shell mass through the
relation (valid to $O(\Gamma_W^2)$)
\begin{equation} \label{eq:on-shell/pole}
\hat{M}_W = M_W+\frac{\Gamma_W^2}{2 M_W}\,,
\end{equation}
where
\begin{equation} \label{eq:mass_width}
\Gamma_W=\frac{3}{4} \frac{\alpha}{s_w^2} M_W=
\frac{3 G_\mu M_W^3}{2\sqrt 2\pi}\,.
\end{equation}
We use the fine-structure constant in the $G_\mu$ scheme,
$\alpha \equiv \sqrt 2 G_\mu M_W^2 s_w^2/\pi$,
and the on-shell Weinberg angle $c_w=M_W/M_Z$. Inserting
(\ref{eq:mass_width}) into (\ref{eq:on-shell/pole}), and solving
the equation for $M_W$, we get the following pole parameters:
\begin{equation}
\begin{aligned}
M_W= 80.377\,\text{GeV} ,
\hspace{4 mm}&
\Gamma_W=2.04483\,
\text{GeV} .
\end{aligned}
\end{equation}
The value of the $W$ width used here is the leading-order decay
width~\eqref{eq:gamma0}, excluding the one-loop QCD correction.
This is appropriate for a tree-level calculation
and ensures that the branching ratios add up to one. Correspondingly we
set $\Delta^{(2)}=0$ in the effective-theory calculation.
In Figure~\ref{fig:eft} we plot the numerical
result obtained with Whizard~\cite{Kilian:2001qz} for the
tree-level cross-section, and the successive effective-theory
approximations. We used the fixed-width scheme in Whizard and
checked that the results from the O'Mega~\cite{Moretti:2001zz},
 CompHep~\cite{Pukhov:1999gg}
and MadGraph~\cite{Stelzer:1994ta} matrix elements agree within the numerical
error of the Monte-Carlo integration.
The large constant shift of about 100~fb
by the N$^{1/2}$LO correction from the
hard region is clearly visible, but the NLO approximation is already
close to the full Born calculation. In Table~\ref{tab:efts} we perform
a more detailed numerical comparison, now including also
the N$^{3/2}$LO approximation. (The
missing energy-independent N$^{3/2}$LO terms are set to zero.)
We observe that the convergence of
the expansion is very good close to the threshold at
$\sqrt{s}\approx 161\,$GeV, as should
be expected. The accuracy of the approximation degrades
as one moves away from threshold, particularly below threshold,
where the doubly-resonant potential configurations are
kinematically suppressed. If one aims at $0.5\%$ accuracy of the
cross section, the NLO approximation suffices only in a rather narrow
region around threshold. Including the N$^{3/2}$LO term
from the first correction in the expansion in the hard region
leads to a clear improvement both
above ($\sim 0.1 \%$ at 170 GeV) and below threshold ($\sim 10 \%$
at 155 GeV). The energy region where the target accuracy is met
now covers the region of interest for the $W$ mass determination 
(see Section~\ref{sec:uncertainty}).

\begin{table}[t]
\begin{center}
\begin{tabular}{|c|c|c|c|c|c|}
\hline&
\multicolumn{4}{c}{ $\sigma(e^-e^+\to \mu^-\bar\nu_\mu u\bar d\,)$(fb)}&
\\\hline
$\sqrt{s}\,[\mbox{GeV}]$ & EFT(LO) & EFT($\sqrt{\mbox{N}}$LO)& EFT(NLO) &
EFT($\mbox{N}^{\frac{3}{2}}$LO) &
exact Born \\
\hline
155 & 101.61 & 1.62 & 43.28 & 31.30 &
34.43(1) \\
\hline
158 & 135.43 & 39.23 & 67.78 &  62.50  &
63.39(2) \\
\hline
161 & 240.85 & 148.44 & 160.45 & 160.89 &
160.62(6) \\
\hline
164 & 406.8 & 318.1 & 313.5 & 318.8  &
318.3(1) \\
\hline
167 & 527.8 & 442.7 & 420.4 &  429.7  &
428.6(2)\\
\hline
170 & 615.5 & 533.9 & 492.9 & 505.4 &
505.1(2)\\
\hline
\end{tabular}
\end{center}
\caption{Comparison of the  numerical computation of the
  full Born result with Whizard  with
successive effective-theory approximations.}
\label{tab:efts}
\end{table}

\section{Radiative corrections}
\label{sec:radcor}

In this section we calculate the NLO contributions that correspond
to genuine loop corrections to four-fermion production. As outlined
in Section~\ref{sec:nlo} there are several such contributions:
an electroweak correction to the matching coefficient of the leading
$W$ pair-production operator and to $W$ decay; a correction from
potential photons associated with the Coulomb force between the
slowly moving $W$ bosons; and soft and collinear photon effects.

\subsection{Hard corrections to  production and decay}
\label{subsec:hard}

The two hard electroweak corrections required for a NLO calculation
are the one-loop corrections $C_{p,LR}^{(1)}$ and $C_{p,RL}^{(1)}$ in the
production operator (\ref{LPNLO}) and the two-loop electroweak
$W$ self-energy $\Delta^{(2)}$, see (\ref{eq:DeltaNLO}).
We reiterate that these are conventional perturbative calculations
performed in a strict expansion in
$\alpha_{ew}$. In particular, in the 't Hooft-Feynman gauge, the
propagators of the massive gauge bosons are simply given by $-i
g^{\mu\nu}/(k^2-M^2)$ and the self-energy insertions are taken into
account perturbatively.  All fermions except for the top quark are
treated as massless.

Before addressing these two calculations separately, we briefly
discuss the renormalization conventions for the parameters and fields of
the electroweak standard model (SM). For a scattering amplitude,
whose tree-level expression is proportional to $g_{ew}^n = (4\pi
\alpha_{ew})^{n/2} = (4\pi \alpha/s_w^2)^{n/2}$ the
one-loop counterterm is given by
\begin{equation}
\label{eq:ctterm}
{\rm [tree]} \left(-n\, \frac{\delta s_w}{s_w} + n\, \delta Z_e +
\frac{1}{2} \sum_{\rm ext} \delta Z_{\rm ext} \right) ,
\end{equation}
where the sum extends over all external lines. As specified
in (\ref{inputs}) the three independent parameters of the
electroweak SM are taken to be the $W$ and $Z$ boson mass,
and the Fermi constant $G_\mu$ (including the electromagnetic
correction to muon decay in the Fermi theory), while
$c_w\equiv M_W/M_Z$ and $\alpha \equiv \alpha_{ew} s_w^2
\equiv \sqrt 2 G_\mu M_W^2 s_w^2/\pi$ are derived quantities.
Similar to the $\alpha(M_Z)$ scheme, the $G_\mu$-scheme
for defining the electromagnetic coupling has the advantage
that the light-fermion masses can be set to zero~\cite{Dittmaier:2001ay,
Denner:1991kt}. The counterterm for $s_w$ is related to
the $W$- and $Z$-boson self-energies. In the $G_\mu$ scheme we
have
\begin{equation}
- \frac{\delta s_w}{s_w} + \delta Z_e
= \frac{1}{s_w c_w} \frac{\Pi^{AZ}_{T}(0)}{M_Z^2} +
\frac{\Pi^{W}_{T}(0)- {\rm Re}\, \Pi^{W}_{T}(M_W^2)}{2 M_W^2}
- \frac{\delta r}{2},
\label{eq:cttermGmu}
\end{equation}
where $\Pi^{W}_{T}$ is the transverse self-energy of the $W$
boson\footnote{In the
conventions used here and in~\cite{Beneke:2004km} the sum of the
amputated 1PI graphs is given by $(-i \Pi)$ which is the opposite sign
compared to~\cite{Denner:1991kt}.}
and
\begin{equation}
\delta r = \frac{\alpha}{4\pi s_w^2}
    \left(6+\frac{7-4 s_w^2}{2 s_w^2} \ln c_w^2\right) 
\label{eq:drdef}
\end{equation}
appears in the explicit expression for the electroweak correction
to muon decay, $\Delta r$ (see e.g.~\cite{Denner:1991kt}).
For the field-renormalization counterterms $\delta Z_{\rm
ext}$ for the external lines we use the conventional on-shell
scheme for wave-function renormalization~\cite{Denner:1991kt}
in accordance with the choice made in
Section~\ref{subsec:NLOpotential} for the renormalized $W$
propagator. In
particular, for the $W$-boson and fermion wave-function
renormalization we have
\begin{equation}
\label{eq:Zext}
\delta Z_{W} =  {\rm Re}
\frac{\partial \Pi^{W}_{T}(p^2)}{\partial p^2}\big|_{p^2=M_W^2} ,
\qquad
\delta Z_{f} =  {\rm Re}\, \Pi^{f}(0) ,
\end{equation}
where
$\Pi^{f}$ denotes the self energy of the fermion. (Note that ${\rm
Re}\, \Pi^{f}(0) = \Pi^{f}(0)$.) The on-shell field renormalization
of the fermions ensures that no further finite renormalization
is needed in calculating the scattering amplitude. On the other
hand, since we never consider a physical process with external
$W$ bosons, the renormalization factor for the $W$ field
is purely conventional, and our final result is
independent of the convention for $\delta Z_{W}$. However, the
matching coefficient of the production operator calculated below
does depend on this convention. The dependence is cancelled by
the dependence of (\ref{eq:NLO-prop}) on $\Pi^{(1,1)}$,
the on-shell derivative of the renormalized one-loop self-energy,
whose value depends on $\delta Z_{W}$.

\subsubsection{Production vertices}
\label{subsubsec:production_hard}

The general method on how to obtain the matching equations needed
to determine the short-distance coefficients of production
operators has been discussed in~\cite{Beneke:2004km}. For
$C_{p,LR}^{(1)}$ and $C_{p,RL}^{(1)}$ we compute the $e^-_{L/R}
\,e^+_{R/L} \to W^- W^+$ scattering amplitude at leading order in
the non-relativistic approximation using dimensional
regularization in $d=4-2\epsilon$ dimensions. This is compared to
the amplitude obtained with the tree-level operator in the
effective theory and the matching coefficient is determined to
make the results agree. The matching coefficients thus determined
are gauge invariant by construction provided the scattering
amplitude is calculated with the external $W$ boson momenta at the
complex pole position. The matching prescription also includes an
additional factor $\sqrt{2
M_W}\,\varpi^{-1/2}$~\cite{Beneke:2004km}, as given in
(\ref{eq:Phinorm}), for each external $\Omega$ field. However,
here we depart from the ``correct'' matching procedure and omit
the factor $\varpi^{-1/2}$, since it was already included in
Section~\ref{subsec:NLOpotential} (see discussion after
(\ref{eq:NLO-prop})).

\begin{figure}[t]
  \begin{center}
  \includegraphics[width=0.8\textwidth]{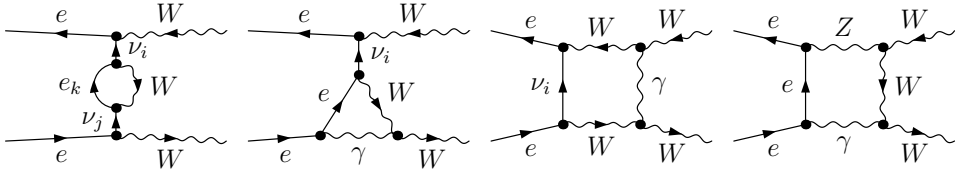}
  \caption{Sample diagrams contributing to the matching
  of the production operator ${\cal O}_p$ at one loop.}
  \label{fig:prodhard}
  \end{center}
\end{figure}

The diagrams for the $e^-(p_1)
e^+(p_2) \to W^-(k_1) W^+(k_2)$ scattering process are generated with
FeynArts~\cite{Hahn:2000kx} and the algebra is performed with
FeynCalc~\cite{Mertig:1990an}.  At one loop, there are 65 two-point
diagrams, 84 three-point diagrams and 31 four-point diagrams
(generically counting up-type quarks, down type quarks, leptons and
neutrinos), some of which are shown in Figure~\ref{fig:prodhard}.
Due to the simplified kinematics, many of these diagrams do
not contribute.  In fact, since the one-loop contributions are already
suppressed by $\alpha_{ew} \sim \delta$ it is sufficient to take the
leading order in the non-relativistic expansion of the one-loop
diagrams and to set $k_1^2$, $k_2^2$ to $M_W^2$ rather than to the complex
pole position. Thus, for the $W$ momenta we use $k_1 = k_2 =
M_W v$ whereas the incoming lepton momenta can be parametrized as
$p_1=(M_W,\vec{p}\,)$ and $p_2=(M_W,-\vec{p}\,)$ with $|\vec{p}\,| =
M_W$. This results in two simplifications. First, many diagrams vanish
consistent with the fact that the tree-level $s$-channel diagrams do
not contribute at leading order in the non-relativistic expansion.
Second, the number of scales present in the loop integrals is
reduced. Due to the simplified kinematics, all box integrals can be
reduced to triangle diagrams and the one-loop correction to the
amplitude for the process $e^-_L e^+_R \to W^- W^+$ takes the simple form
\begin{equation}
\label{eq:ampform}
{\cal A}_{WW} =
 \frac{\pi\alpha_{ew}}{M_W^2} \,C_{p,LR}^{(1)}\,(p_1-p_2)_\mu\,
  \langle p_2-|\!\not{\!\eps_3}\eps_4^\mu+
  \!\not{\!\eps_4}\eps_3^\mu\,|p_1-\rangle
\end{equation}
expected from~(\ref{LPNLO}), with $\eps_{3,4}^\mu$ denoting the
polarization vectors of the $W$ bosons. (For $h=RL$, the fermion
helicities are reversed.) The scalar coefficients
$C_{p,h}^{(1)}$ can be obtained by projections of the full
amplitude. Thus, we are left with the calculation of a scalar quantity
and standard techniques for the reduction of tensor and scalar
integrals can be applied.

In the computation of $C_{p,RL}^{(1)}$ all poles cancel and we are
left with a finite result. This is to be expected, since the
corresponding Born term vanishes, as indicated in~(\ref{LPlead}). For
$C_{p,LR}^{(1)}$, the matching coefficient of the operator that does
not vanish at tree level, the poles do not cancel. After adding the
counterterm~(\ref{eq:ctterm}) with $n=2$, it takes the form
\begin{equation}
C_{p,LR}^{(1)} = \frac{\alpha}{2\pi} \left[
   \left(-\frac{1}{\epsilon^2} - \frac{3}{2\epsilon}\right)
   \left(-\frac{4 M_W^2}{\mu^2}\right)^{\!-\epsilon}
   + c_{p,LR}^{(1,{\rm fin})} \right] ,
\label{eq:CLR}
\end{equation}
where the finite part $c_{p,LR}^{(1,{\rm fin})}$ together with the
expression for $C_{p,RL}^{(1)}$ is given explicitly in
Appendix~\ref{ap:hard1loop}. For the final expression of the matching
coefficient, the poles have to be subtracted. However, we leave them
explicit in order to demonstrate their cancellation against (double)
poles from the soft contribution and poles related to initial-state
collinear singularities. Numerically,
\begin{equation}
c_{p,LR}^{(1,{\rm fin})} =-10.076 + 0.205 i
\end{equation}
for $M_W=80.377\,$GeV, $M_Z=91.188\,$GeV, top-quark mass
$m_t=174.2\,$GeV and Higgs mass $M_H=115\,$GeV.

The matching coefficients $C_{p,LR}^{(1)}$ and $C_{p,RL}^{(1)}$
both have a non-vanishing imaginary part. Taken at
face value, this imaginary part contributes to the imaginary part
of the forward scattering amplitude ${\cal A}$ and, therefore, to the
total cross section.  Denoting by ${\cal A}_{\Delta C}^{(1)}$ the NLO
contribution to ${\cal A}$ resulting from $C_{p}^{(1)}$ we have
\begin{equation}
\label{eq:ImAC}
{\rm Im}\,  {\cal A}_{\Delta C}^{(1)} =
{\rm Im} \left(2 C_{p}^{(1)}{\cal A}^{(0)}\right) =
2\,  {\rm Re}\, C_{p}^{(1)}\ {\rm Im}\, {\cal A}^{(0)} +
2\, {\rm Im}\, C_{p}^{(1)}\ {\rm Re}\, {\cal A}^{(0)} .
\end{equation}
However, the second term in (\ref{eq:ImAC}) is induced by cuts that do
not correspond to the final state we are interested in, such as the
$Z\gamma$ intermediate state in the fourth diagram of
Figure~\ref{fig:prodhard}. In fact, at leading order in the
non-relativistic expansion, none of the diagrams that contribute to
the hard matching coefficients contains either a quark or a muon. To
obtain the flavour-specific cross section we are concerned with, we
therefore have to drop the second term in (\ref{eq:ImAC}) and in what
follows it is always understood that we take the real part of the
matching coefficients $C_{p,LR}^{(1)}$ and $C_{p,RL}^{(1)}$. Recalling
the discussion of cut (2) at the end of
Section~\ref{subsec:NLOpotential}, we note that beyond NLO the
situation is more complicated, as some of the cuts contributing to the
imaginary part of the matching coefficient $C_{p}$ do correspond to
the flavour-specific cross section we are interested in.

The contribution to the cross section resulting from the NLO correction to
the production operators is obtained by multiplying the
imaginary part of
 $ {\cal A}_{\Delta C}^{(1)}$  by the leading order branching ratios.
The correction to the cross section for the $e_L^-e_R^+$ polarization
is therefore given by
\begin{equation}
  \Delta \sigma^{(1)}_{\text{hard}}=
  \frac{1}{27s } \; 2\, \re \,C_{p,LR}^{(1)} \; \im \,{\cal
    A}_{LR}^{(0)}.
\label{shard}
\end{equation}
Because there is no interference of the  helicities
 $e_R^-e_L^+$ and $e_L^-e_R^+$, the
coefficient $C_{p,RL}^{(1)}$ does not contribute at NLO.
Introducing the abbreviations
\begin{equation}
 \eta_{-}=r^0-\frac{\vec{r}^{\,2}}{2 M_W}+i\frac{\Gamma^{(0)}_W}{2},
\qquad
 \eta_{+}=E-r^0-\frac{\vec{r}^{\,2}}{2
 M_W}+i \frac{\Gamma^{(0)}_W}{2}
\end{equation}
for the non-relativistic propagators in the leading-order
diagram, Figure~\ref{fig:LO}, and $\tilde \mu^2=\mu^2
e^{\gamma_E}/(4 \pi)$, we can rewrite (\ref{shard}) as
\begin{eqnarray}
 \Delta \sigma^{(1)}_{\text{hard}}
 &=& \frac{16\pi^2\alpha^2_{ew}}{27 M_W^2 s} \,
  \mbox{Im}\,\bigg\{(-i) \tilde\mu^{2\epsilon}\!
\int \frac{d^d r}{(2 \pi)^d} \,\frac{1-\epsilon}{\eta_{-} \eta_{+}} \bigg\}
\nonumber\\
&& \times \,
 2\,\re\, \frac{\alpha}{2\pi}\left[
   \left(-\frac{1}{\epsilon^2} - \frac{3}{2\epsilon}\right)
   \left(-\frac{4 M_W^2}{\mu^2}\right)^{-\epsilon}
   + c_{p,LR}^{(1,{\rm fin})} \right].
\label{eq:hardsigma}
\end{eqnarray}
The unintegrated  form of the result is given to
make the cancellation of the $\epsilon$-poles against other contributions
computed in the following subsections more transparent.

\subsubsection{Decay corrections}
\label{subsubsec:decay_hard}

Next we discuss the electroweak correction to the matching coefficient
$\Delta$. In the pole mass and on-shell field renormalization scheme
$\Delta^{(2,ew)} = -i \Gamma^{(1,ew)} = i M_W
\,\mbox{Im}\,\Pi^{(2,0)}$. The cuts of the 2-loop electroweak $W$
self-energy consist of two parts, corresponding to the virtual and
real hard corrections to the $W$ pole decay width. Dealing with the
total cross section, we only need the sum of these two. However, we
also have to discuss how to obtain results for the flavour-specific
process $e^+ e^-\to\mu^- \bar{\nu}_\mu\, u\, \bar{d} \, X$. To aid
this, we will discuss the virtual and real corrections separately,
starting with the former.

\begin{figure}[t]
  \begin{center}
  \includegraphics[width=0.7\textwidth]{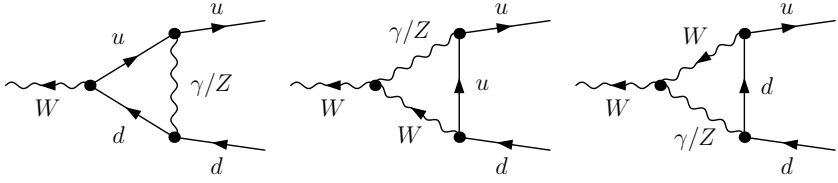}
  \caption{Diagrams contributing to the virtual correction $C_{d,
  h}^{(1)}$ at one loop.}
  \label{fig:decayhard}
  \end{center}
\end{figure}

The virtual one-loop correction to the pole-scheme decay width into a
single lepton $(l)$ or quark $(h)$ doublet can be written as
\begin{equation}
\Gamma_{W, l/h}^{(1,{\rm virt})} =
 2\, \Gamma_{W, l/h}^{(0)}\,
{\rm Re}\, C_{d, l/h}^{(1)} ,
\label{eq:GammaVirt}
\end{equation}
where the tree-level widths in $d$ dimensions are $\Gamma_{W,
l}^{(0)} = \Gamma^{(0)}_{\mu^-\bar\nu_\mu}
=\alpha_{ew} M_W/12 + {\cal O}(\epsilon)$ and $\Gamma_{W,
h}^{(0)} = \Gamma^{(0)}_{u\bar d} = 3\, \Gamma_{W,l}^{(0)}$.
The calculation of $C_{d, h}^{(1)}$ involves the
evaluation of the diagrams depicted in Figure~\ref{fig:decayhard} with
obvious modifications for the leptonic decay. After adding the
counterterm (\ref{eq:ctterm}) with $n=1$ we obtain
\begin{equation}
\label{eq:decayvirtual}
C_{d, l/h}^{(1)} =  \frac{\alpha}{2\pi} \left[
\left(-\frac{1}{2 \epsilon^2} - \frac{5}{4\epsilon}\right)
    \left(\frac{M_W^2}{\mu^2}\right)^{\!-\epsilon}
+ Q_f\bar{Q}_f \left(- \frac{1}{\epsilon^2}
                  - \frac{3}{2\epsilon}\right)
  \left(-\frac{M_W^2}{\mu^2}\right)^{\!-\epsilon}
  + c_{d, l/h}^{(1,{\rm fin})} \right],
\end{equation}
where for the leptonic (hadronic) decay we have to set the electric
charges to $Q_f=-1, \bar{Q}_f=0$ ($Q_f=2/3, \bar{Q}_f=-1/3$ ). The
finite parts $c_{d, l/h}^{(1,{\rm fin})}$ of the matching coefficients
are given explicitly in Appendix~\ref{ap:hard1loop}.
Numerically,
\begin{equation}
c_{d,l}^{(1,{\rm fin})} = -2.709-0.552\,i,
\qquad
c_{d,h}^{(1,{\rm fin})} = -2.034-0.597\,i,
\end{equation}
for $M_W=80.377\,$GeV, $M_Z=91.188\,$GeV,
$m_t=174.2\,$GeV, and $M_H=115\,$GeV.

To this we have to add the correction due to hard real radiation
of a single photon. Since the corresponding soft corrections vanish,
the hard real corrections are equivalent to the real
corrections evaluated in the standard electroweak theory and
their calculation is straightforward. We compute the bremsstrahlung
diagrams and integrate the squared amplitude (divided by $2M_W$) over
the $d$-dimensional phase-space~\cite{Marciano:1975de}. The expression
thus obtained contains infrared (double) poles which cancel the poles
in (\ref{eq:GammaVirt}) and we are left with finite expressions for
the flavour-specific leptonic and hadronic matching coefficients.
Including the (two-loop) QCD correction to the hadronic decay, they
read
\begin{eqnarray}
\label{eq:Deltal}
\Delta^{(2)}_{l} &=&
-i\, \Gamma^{(1,ew)}_{W,l},
\nonumber \\
\label{eq:Deltah}
\Delta^{(2)}_{h} &=&
-i\left[\Gamma^{(1,ew)}_{W,h}+ 1.409
  \,\frac{\alpha_s^2}{\pi^2}\, \Gamma_{W,h}^{(0)}\right] ,
\nonumber \\
\label{eq:Gamma1ewFS}
\Gamma^{(1,ew)}_{W,l/h} &=&
\Gamma_{W,l/h}^{(0)} \,
\frac{\alpha}{2\pi}\, \left[ 2\, {\rm Re}\, c_{d, l/h}^{(1,{\rm fin})} +
\left( \frac{101}{12} +
\frac{19}{2} Q_f \bar{Q}_f - \frac{7\pi^2}{12}
- \frac{\pi^2}{6} Q_f \bar{Q}_f \right) \right].
\end{eqnarray}
Strictly speaking, for the computation of these matching coefficients we
have to expand around the complex pole $\bar{s}$ and not around
$M_W^2$. However, the difference in the width is of order $\alpha^3$
and thus beyond NLO~\cite{Sirlin:1991fd}.

\subsection{Coulomb corrections}
\label{subsec:coulomb}

\begin{figure}[t]
  \begin{center}
  \includegraphics[width=0.40\textwidth]{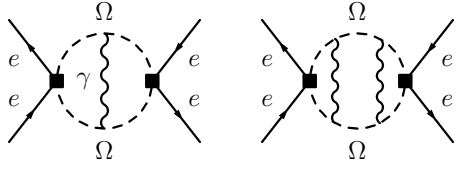}
\caption{First and second order Coulomb correction.}
\label{fig:coulomb}
  \end{center}
\end{figure}

The exchange of potential photons with energy $k_0 \sim M_W\delta $
and three-momentum $\vec{k}\sim M_W\sqrt{\delta}$, shown in
Figure~\ref{fig:coulomb}, corresponds to insertions of the
non-local four-boson interactions in the effective
Lagrangian (\ref{LPNR}). These insertions can be summed
to all orders in terms of the Green function $G_c(\vec{r},
\vec{r}^{\,\prime};E)$ of the Schr\"odinger operator $-\vec{\nabla}^2/M_W
-\alpha/r$ evaluated at $\vec{r}=\vec{r}^{\,\prime}=0$.
Using the representation of the Green function
given in~\cite{Wichmann:1961a}, we obtain \cite{Beneke:1999zr}
\begin{equation}
i{\cal A}_{\rm coulomb} = - 4 i\pi\alpha_{ew}^2 \alpha
\left\{\frac{1}{2}\ln \left(-\frac{E+i \Gamma_W^{(0)}}{M_W}\right)
+\psi\!\left(1-\frac{\alpha}{2\sqrt{-(E+ i \Gamma_W^{(0)})/M_W}}
\right)
\right\},
\end{equation}
where $\psi(x)$ is Euler's psi-function, and a subtraction-scheme
dependent real constant that drops out in the cross section has
been omitted.  The diagram with no photon exchange is not 
included in this expression, since it corresponds to the leading-order
amplitude (\ref{eq:Alead}). The logarithm constitutes a
$\alpha/\sqrt{\delta}\sim\sqrt{\delta}$ correction relative to
the leading-order scattering amplitude (\ref{eq:Alead}). The
expansion of the psi-function in $\alpha$ results in an expansion
in powers of $\sqrt{\delta}$. Thus, the Coulomb correction
up to NLO reads
\begin{equation}
\Delta \sigma^{(1)}_{\mbox{\tiny Coulomb}} = \frac{4 \pi \alpha^2}{27
  s_w^4 s}\,
\mbox{Im} \left[
-\frac{\alpha}{2} \ln \left(-\frac{E+i \Gamma_W^{(0)}}{M_W}\right)
+\frac{\alpha^2\pi^2}{12} \,\sqrt{-\frac{M_W}{E+i \Gamma_W^{(0)}}}
\,\,\right].
\label{eq:coulomb}
\end{equation}
This contributes only to the LR helicity cross section, since the
production operator at the vertices in Figure~\ref{fig:coulomb}
is the leading order one (\ref{LPlead}).
Directly at threshold ($E=0$) the one-photon exchange
N$^{1/2}$LO term (the logarithm in (\ref{eq:coulomb}))
is of order 5\% relative to
the leading order. Two-photon exchange is only a few-permille correction,
confirming the expectation that Coulomb exchanges do not have
to be summed to all orders due to the large width of
the $W$ boson. The one and two Coulomb-exchange terms have already
been discussed in~\cite{Fadin:1993kg,Fadin:1995fp}.

\subsection{Soft-photon corrections}
\label{subsec:soft}

\begin{figure}[t]
  \begin{center}
  \includegraphics[width=0.7\textwidth]{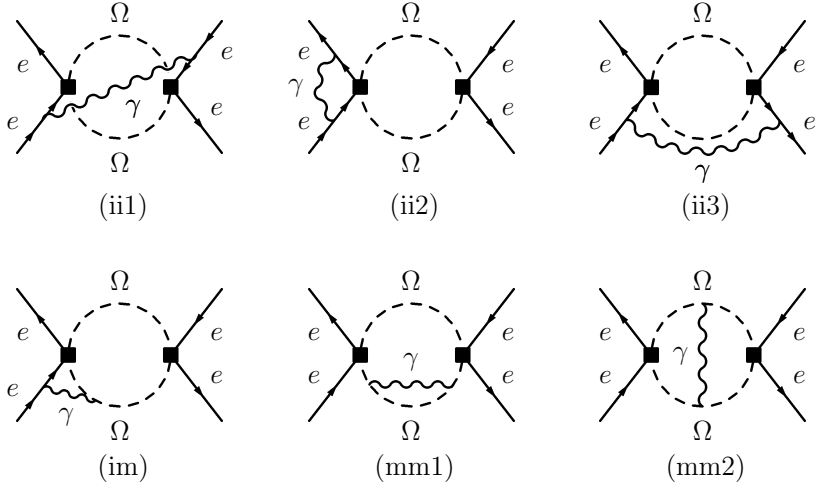}
\caption{Soft-photon diagrams in the effective theory:
Initial-initial state interference (ii),
initial-intermediate state interference (im) and
intermediate-intermediate state interference (mm).
Symmetric diagrams are not shown.}
\label{fig:soft}
  \end{center}
\end{figure}

We now turn to the radiative correction originating from soft-photon
exchange.  These are $O(\alpha)$ contributions to
the forward-scattering amplitude, and correspond to two-loop
diagrams in the effective theory
containing a photon with momentum components $q_0 \sim
|\vec{q}\,| \sim M_W \delta$.  The relevant Feynman rules are given by
the coupling of the soft photon to the $\Omega_\pm $ fields in the PNRQED
Lagrangian~\eqref{LPNR} and to the collinear electrons and positrons
contained in the SCET Lagrangian.  The latter is simply the eikonal
coupling $\pm i e n^\mu$, where $n^\mu$ is the direction of the
four-momentum of the electron or positron. The topologies contributing to the
two-loop forward-scattering amplitude are shown in
Figure~\ref{fig:soft}. The $W$-boson vertices are leading-order
production vertices, hence at NLO the soft correction applies only
to the left-right $e^- e^+$ helicity forward-scattering amplitude.
Note that (mm2) is not a double-counting
of the Coulomb-exchange diagram in Figure~\ref{fig:coulomb}, since
the two diagrams refer to different loop momentum regions.

It is well known that for the process $e^-e^+\to W^- W^+\to f_1\bar
f_2 f_3\bar f_4$ the soft-photon corrections related to the final
state cancel for the inclusive cross
section~\cite{Fadin:1993dz,Melnikov:1993np}.
The diagrams of type (im) in Figure~\ref{fig:soft}
cancel pairwise when the sum over incoming positrons and electrons
is performed.
The sum of the diagrams of the form of (mm1) and (mm2)
 cancels after the loop integrals are performed.
Therefore the sum of all diagrams where a soft photon couples to
an $\Omega$ line vanishes.
In the effective theory this cancellation can be seen from the outset,
since it follows from the particular form of the leading coupling
of a soft photon to non-relativistic $W$ bosons in the effective
Lagrangian~\eqref{LPNR}, which involves only $A_s^0(t,0)$.
Since the residual gauge invariance of the
effective Lagrangian allows one to set the time-like component of the
photon field to zero, at leading order the $\gamma \Omega_{\mp}
\Omega_{\mp}$ couplings can be removed from the Lagrangian.

Therefore the soft-photon correction in the
effective theory is given by the initial-initial state
interference diagrams. However, diagram (ii2) leads to a scaleless
integral which vanishes in dimensional regularization, and diagram
(ii3) and the symmetric diagram are proportional to $p_1^2 \sim 0$ and
$p_2^2\sim 0$, respectively. The only non-zero diagram is (ii1)
and the corresponding crossed diagram. The sum of the
two diagrams evaluates to
\begin{eqnarray}\label{eq:finalsoft}
\Delta \mathcal{A}_{\text{soft}}^{(1)}&=&
 \frac{16\pi^2 \alpha_{ew}^2} { M_W^2}
\,8\pi\alpha \,(p_1\cdot p_2)\,(1-\epsilon)\,
\tilde\mu^{4 \epsilon} \!
\int \frac{d^d r}{(2 \pi)^d} \int \frac{d^d q}{(2 \pi)^d} \nonumber\\
&&\times \frac{1}{\eta_{+}} \, \frac{1}{(q^2 +i \epsilon)}\,
\frac{1}{(-q \cdot p_1+i \epsilon)}\, \frac{1}{(-q \cdot p_2+i \epsilon)}\,
\frac{1}{(\eta_--q_0)}\nonumber\\
&=& \frac{16\pi^2\alpha^2_{ew}}{M_W^2} \,\frac{\alpha}{ \pi}\,
(-i)\,\tilde\mu^{2 \epsilon}\!
\int \frac{d^d r}{(2 \pi)^d} \,\frac{1-\epsilon}{\eta_{-} \eta_{+}}
\nonumber\\
&&\times \left[\frac{1}{\epsilon^2}
-\frac{2}{\epsilon} \ln \left(-\frac{2 \eta_{-}}{\mu}\right)
+2 \ln^2 \left(-\frac{2 \eta_{-}}{\mu}\right)+\frac{5\pi^2}{12} \,
\right].
\end{eqnarray}
The double $\epsilon$-pole in (\ref{eq:finalsoft}) cancels
against the pole in the hard matching coefficient; the single
pole  can be factorized into the initial-state electron (positron)
structure function as shown in Section~\ref{sec:isr}.
Subtracting the pole part of the integrand (\ref{eq:finalsoft})
before performing the integration, one obtains
\begin{equation}
\Delta \mathcal{A}^{(1,\text{fin})}_{\text{soft}}= \mathcal{A}^{(0)}_{LR}
 \,\frac{2\alpha}{\pi}
\left[\ln^2 \left(-\frac{8 (E+i
\Gamma_W^{(0)})}{\mu}\right)-4 \ln
\left(-\frac{8 (E+i \Gamma_W^{(0)})}{\mu}\right)+8+\frac{13}{24}
\pi^2\right].
\end{equation}
As before, the $r^0$ integration has been
performed by closing the $r^0$ integration contour in the upper half-plane
and picking up the pole at $r^0=E-\vec r^{\,2}/(2
M_W)+i\Gamma_W^{(0)}/2$.
Because of the absence of soft corrections related to the final state,
at NLO the soft corrections to the flavour-specific
process~\eqref{eq:wwprocess} can be obtained by multiplying the soft
two-loop contributions to the forward-scattering amplitude by the
leading-order branching ratios, thus
\begin{equation}
\Delta \sigma^{(1)}_{\mbox{\tiny soft}} = \frac{1}{27 s}
\,\mbox{Im}\,\Delta \mathcal{A}_{\text{soft}}^{(1)}.
\label{ssoft}
\end{equation}
As a check, we also calculated the soft
corrections directly for the process~\eqref{eq:wwprocess} and found
agreement with the simpler calculation of the forward-scattering
amplitude.

\subsection{Collinear-photon corrections}

\begin{figure}[t]
  \begin{center}
  \includegraphics[width=0.4\textwidth]{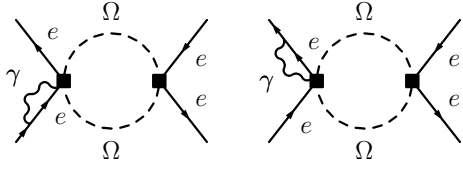}
\caption{Collinear-photon diagrams in the effective theory.
Two symmetric diagrams are not shown.}
\label{fig:collinear}
  \end{center}
\end{figure}

Finally we consider collinear-photon corrections, corresponding
to photon energies of order $M_W$, and photon virtuality of order
$M_W\Gamma_W$. The four-momentum of the photon is proportional
to the initial-state electron or positron momentum. The collinear
photon couplings arise from the SCET Lagrangian, while their
couplings to the $W$ bosons is encoded in the collinear Wilson lines
in the production operators. The diagrams corresponding to
NLO contributions are shown in Figure~\ref{fig:collinear}. As
discussed in \cite{Beneke:2004km} all these diagrams are scaleless
for on-shell, massless initial-state particles. However, we shall
have to say more about collinear effects in Section~\ref{sec:isr},
when we include the resummation of large initial-state radiation
logarithms.

\subsection{Summary of radiative corrections}

The radiative correction to the next-to-leading order cross section
is given by the sum of the corrections (\ref{eq:hardsigma}),
(\ref{ssoft}), (\ref{eq:coulomb}), (\ref{eq:decaycorr})
computed in the previous sections,
\begin{equation}
\label{eq:nlototal}
\hat{\sigma}^{(1)}_{LR}=\Delta \sigma^{(1)}_{\mbox{\tiny hard}}
+\Delta \sigma^{(1)}_{\mbox{\tiny soft}}
+\Delta \sigma^{(1)}_{\mbox{\tiny Coulomb}}
+\Delta \sigma^{(1)}_{\mbox{\tiny decay}}\,.
\end{equation}
Recall that this refers to the $e_L^- e_R^+$ helicity initial state, while
there are no radiative corrections to the other helicity combinations
at NLO. The radiative correction to the
unpolarized cross section is one fourth of the LR contribution.

Because of the approximation $m_e=0$, the cross section
is not infrared-safe, as can be seen by summing the
four contributions. The Coulomb and
decay corrections are free of infrared singularities.
For the sum of the soft (\ref{eq:finalsoft}) and hard
(\ref{eq:hardsigma}) terms we obtain the following expression:
\begin{eqnarray}\label{eq:hard+soft}
\Delta \sigma^{(1)}_{\mbox{\tiny hard}}
+\Delta \sigma^{(1)}_{\mbox{\tiny soft}}&=&
\frac{16\pi^2\alpha^2_{ew}}{27 M_W^2 s} \,\frac{\alpha}{ \pi}
\,\mbox{Im} \,\Bigg\{ (-i ) \,\tilde\mu^{2\epsilon}\!
\int \frac{d^d r}{(2 \pi)^d} \,\frac{1-\epsilon}{\eta_{-} \eta_{+}}
\nonumber\\
&&\times \left[-\frac{1}{\epsilon}
\left(2 \ln \left(-\frac{\eta_{-}}{M_W}\right)
+\frac{3}{2}\right)+2 \ln^2 \left(-\frac{2 \eta_{-}}{\mu}\right)
-2 \ln^2 \left(\frac{2 M_W}{\mu}\right)\right.
\nonumber\\
&&\left.\hspace{0.6cm}
+\,3 \ln \left(\frac{2 M_W}{\mu}\right)+
\mbox{Re}\,\Big[c_{p,LR}^{(1,\rm fin)}\Big]
+\frac{11\pi^2}{12} \,\right]\Bigg\}\,.
\end{eqnarray}
The cross section $\hat{\sigma}^{(1)}_{LR}$ is a ``partonic'' cross
section. It should be convoluted with the electron (positron)
distribution function, which contains the infrared effects
associated with the electron mass scale. In the following section we
discuss how the partonic cross section is transformed to the
infrared-finite physical cross section.

\section{Initial-state radiation}
\label{sec:isr}

The remaining $\epsilon$-poles in (\ref{eq:hard+soft}) are associated
with emission of photons collinear to the incoming electron or
positron, and can be
factorized into the electron distribution function
$\Gamma_{ee}^{\overline{\mbox{\tiny MS}}}$, in terms of which the physical
cross section $\sigma$ reads~\cite{Kuraev:1985hb,Beenakker:1994vn}
\begin{equation}
\sigma_h(s) = \int_0^1 dx_1 \int_0^1 dx_2 \,
\Gamma_{ee}^{\overline{\mbox{\tiny MS}}}(x_1)
\Gamma_{ee}^{\overline{\mbox{\tiny MS}}}(x_2)
\,\hat{\sigma}_h^{\overline{\mbox{\tiny MS}}}(x_1 x_2 s)\,.
\end{equation}
Here $\hat{\sigma}_h^{\overline{\mbox{\tiny MS}}}(s)=
\sigma_{h,\mbox{\tiny Born}}(s)+
\hat{\sigma}_{h,\overline{\mbox{\tiny MS}}}^{(1)}(s)$ is our result
for the NLO helicity-specific cross section after adding the
Born cross section from Section~\ref{sec:tree}  and
the radiative correction from  (\ref{eq:nlototal}) with the
 infrared $\epsilon$-poles minimally subtracted. The partonic
cross section depends on the scales $Q=\{M_W, E, \Gamma_W\}$
and the factorization scale $\mu$. The electron distribution function
in the $\overline{\rm MS}$ scheme
depends on $\mu$ and the very-long distance scale $m_e$.
The physical cross section is independent of $\mu$ and includes the
electron-mass dependence up to effects suppressed by powers
of $m_e/Q$. By evolving the electron distribution from the
scale $m_e$ to the scale $Q$, one sums large collinear logarithms
$\alpha^{n_1}\ln^{n_2} \left(Q^2/m_e^2\right)$, with $n_1=1,...,\infty$,
$n_2=1,...,n_1$ from initial-state radiation of photons
to all orders in perturbation theory. A NLO
calculation of the partonic cross section should go along with
a next-to-leading logarithmic approximation, where all terms
with $n_2=n_1$ and $n_2=n_1-1$ are summed. Note that here we do
not attempt to sum logarithms of $M_W/\Gamma_W$, which are
less important, although the effective-theory formalism is ideally
suited for this summation as well.

Unfortunately the structure functions $\Gamma_{ee}^{\mbox{\tiny LL}}(x)$
available in the literature do not correspond to the
$\overline{\rm MS}$ scheme and sum only leading logarithms
$\alpha^{n}\ln^{n} \left(Q^2/m_e^2\right)$. To convert our result
$\hat{\sigma}_h^{\overline{\mbox{\tiny MS}}}(s)$ to this scheme
and sum the leading-logarithmic initial-state radiation effects
we proceed as follows: first, using the expansion
$\Gamma_{ee}^{\overline{\mbox{\tiny MS}}}(x)=
\delta(1-x)+\Gamma_{ee}^{\overline{\mbox{\tiny MS}},(1)}(x)+
O(\alpha^2)$, we compute the scheme-independent NLO physical cross section
without summation of collinear logarithms,
\begin{equation}
\sigma_h^{\rm NLO}(s) =
\sigma_{h,\mbox{\tiny Born}}(s)+
\hat{\sigma}_{h,\overline{\mbox{\tiny MS}}}^{(1)}(s)
+ 2 \int_0^1 dx\,\Gamma_{ee}^{\overline{\mbox{\tiny MS}},(1)}(x)
\,\sigma_{h,\mbox{\tiny Born}}(x s).
\label{nlophys1}
\end{equation}
Then, by comparing this to the corresponding equation in the
conventional scheme,
\begin{equation}
\sigma_h^{\rm NLO}(s) =
\sigma_{h,\mbox{\tiny Born}}(s)+
\hat{\sigma}_{h,\mbox{\tiny conv}}^{(1)}(s)
+ 2 \int_0^1 dx\,\Gamma_{ee}^{\mbox{\tiny LL},(1)}(x)
\,\sigma_{h,\mbox{\tiny Born}}(x s),
\label{nlophys2}
\end{equation}
we determine $\hat{\sigma}_{h,\mbox{\tiny conv}}^{(1)}(s)$, and hence
$\hat{\sigma}_h^{\mbox{\tiny conv}}(s)=
\sigma_{h,\mbox{\tiny Born}}(s)+
\hat{\sigma}_{h,\mbox{\tiny conv}}^{(1)}(s)$. Finally, we
calculate the initial-state radiation resummed cross section
\begin{equation}
\label{eq:physicalcross}
\sigma_h(s) = \int_0^1 dx_1 \int_0^1 dx_2 \,
\Gamma_{ee}^{\mbox{\tiny LL}}(x_1)
\Gamma_{ee}^{\mbox{\tiny LL}}(x_2)
\hat{\sigma}_h^{\mbox{\tiny conv}}(x_1 x_2 s)
\end{equation}
in the conventional scheme for the electron (positron) distribution
functions. Note that since the Born cross section for the RL helicity
combination is already a NLO effect, the scheme conversion must be
performed only for $h=LR$. For $h=RL$ we simply have
$\hat{\sigma}_{RL}^{\mbox{\tiny conv}}(s)
=\hat{\sigma}_{RL}^{\overline{\mbox{\tiny MS}}}(s) =
\sigma_{RL,\mbox{\tiny Born}}(s)$.

\paragraph{\it Step 1: Calculation of the fixed-order physical cross
  section $\sigma_{LR}^{\rm NLO}(s)$.}

Rather than calculating the last term on the right-hand side of
(\ref{nlophys1}), we compute directly the radiative correction
to the physical cross section,
$\sigma_{LR}^{(1)}(s)$, by converting
$\hat{\sigma}_{h,\overline{\mbox{\tiny MS}}}^{(1)}(s)$,
where the collinear divergences are regulated dimensionally,
into the expression when the electron mass itself is used as the
regulator.

In the presence of the new scale $m_e\ll \Gamma_W, E, M_W$ there
are two new momentum regions that give non-zero contributions
to the radiative corrections. They correspond to
\emph{hard-collinear} photon momentum
($q^0\sim M_W$, $q^2\sim m_e^2$) and
\emph{soft-collinear} photons ($q^0\sim
\Gamma_W$, $q^2 \sim m_e^2\,\Gamma_W^2/M_W^2$).\footnote{The existence
  of two collinear momentum regions is related to the fact that
  the $W$ pair-production threshold region probes the electron distribution
  function near $x=1$, where hard-collinear real radiation is
  inhibited.} The
corresponding loop integrals are scaleless when $m_e=0$; for
$m_e\not=0$, they supply the difference
\begin{equation}
\sigma_{LR}^{(1)}(s) - \hat{\sigma}^{(1)}_{LR} =
\Delta \sigma_{\text{s-coll}}^{(1)}+
\Delta \sigma_{\text{h-coll}}^{(1)}.
\end{equation}
In other words $\sigma_{LR}^{(1)}(s)$ is the sum of the four
contributions in (\ref{eq:nlototal}) plus those from the
two new momentum regions.

Only a small subset of all the radiative correction diagrams has
hard- or soft-collinear contributions, namely those containing a photon
line connecting to an external electron or positron. The topology of
the soft-collinear and hard-collinear
diagrams is identical to the (ii) and (im) diagrams in
Figure~\ref{fig:soft}, and to the diagrams in
Figure~\ref{fig:collinear}, respectively. The calculation is
straightforward. In each region we simplify the integrand by
neglecting all small terms, since the leading-order term in
the expansion in each region is sufficient. The soft-collinear
correction is
\begin{eqnarray}
\label{eq:softcoll}
\Delta \sigma_{\text{s-coll}}^{(1)}&=&
\frac{16\pi^2\alpha^2_{ew}}{27 M_W^2 s} \,\frac{\alpha}{\pi}\,
\mbox{Im} \left\{(-i)\,
\tilde\mu^{2\epsilon}\int \frac{d^d r}{(2 \pi)^d}
\,\frac{1-\epsilon}{\eta_{-} \eta_{+}}\right. \nonumber\\
&&\left.\times \left[-\frac{1}{\epsilon^2}+
\frac{2}{\epsilon} \ln\left(-\frac{m_e \eta_{-}}{\mu M_W}\right)
-2 \ln^2 \left(-\frac{m_e \eta_{-} }{\mu M_W}\right)-
\frac{3\pi^2}{4}\right]\right\}\,,
\end{eqnarray}
the hard-collinear correction
\begin{eqnarray}
\label{eq:hardcoll}
\Delta \sigma_{\text{h-coll}}^{(1)}&=&
\frac{16\pi^2\alpha^2_{ew}}{27 M_W^2 s} \,\frac{\alpha}{ \pi}\,
\mbox{Im} \left\{(-i) \,
\tilde\mu^{2\epsilon}\int \frac{d^d r}{(2 \pi)^d}
\,\frac{1-\epsilon}{\eta_{-} \eta_{+}}\right. \nonumber\\
&&\hspace{-1 cm}\left.\times \left[\frac{1}{\epsilon^2}+
\frac{1}{\epsilon}\left[-2 \ln\left(\frac{m_e}{\mu}\right)
+\frac{3}{2}\right]+2 \ln^2 \left(\frac{m_e}{\mu}\right) -
3 \ln\left(\frac{m_e}{\mu}\right)+\frac{\pi^2}{12}+3\right]\right\}\,.
\end{eqnarray}
The structure of the logarithms makes it clear that the two
contributions arise each from a single scale, $\mu\sim
m_e\Gamma_W/M_W$ and $\mu\sim m_e$, respectively. Adding
(\ref{eq:nlototal}), (\ref{eq:softcoll}), (\ref{eq:hardcoll}), and
making use of (\ref{eq:hard+soft}) results in the
factorization-scheme independent radiative correction to
the physical cross section,
\begin{eqnarray}
\sigma^{(1)}_{LR}(s)
& =& \frac{16\pi^2\alpha^2_{ew}}{27 M_W^2 s} \,\frac{\alpha}{ \pi}\,
\mbox{Im} \,\Bigg\{ (-i) \, \tilde\mu^{2\epsilon}
\int \frac{d^d r}{(2 \pi)^d} \frac{1}{\eta_{-} \eta_{+}}
\left[4 \ln\left(-\frac{\eta_-}{M_W}\right) \ln\left(\frac{2 M_W}{m_e}\right)
\right.
\nonumber\\
&&\left. + \,3 \ln \left(\frac{2 M_W}{m_e}\right)
+\mbox{Re}\,\Big[c_{p,LR}^{(1,\rm fin)}\Big]
+\frac{\pi^2}{4}+3\right]\Bigg\}
+\Delta \sigma^{(1)}_{\mbox{\tiny Coulomb}}
+\Delta \sigma^{(1)}_{\mbox{\tiny decay}}
\nonumber\\
&=&
\frac{4\alpha^3}{27 s_w^4 s}\,\im\,
\Bigg\{(-1) \,\sqrt{-\frac{E+i \Gamma^{(0)}_W}{M_W}}\,
\bigg(  4\ln\bigg(-\frac{4(E+i\Gamma^{(0)}_W)}{M_W}\bigg)
 \ln \left(\frac{2 M_W}{m_e}\right)
\nonumber \\
&&-\,5 \ln \left(\frac{2 M_W}{m_e}\right)
+ \mbox{Re}\,\Big[c_{p,LR}^{(1,\rm fin)}\Big]
+\frac{\pi^2}{4}+3\bigg)\Bigg\}
+\Delta \sigma^{(1)}_{\mbox{\tiny Coulomb}}
+\Delta \sigma^{(1)}_{\mbox{\tiny decay}}\,.
\label{eq:totcollcross}
\end{eqnarray}
After performing the $r$-integral we may set $d$ to four and obtain
a finite result. As expected the $\epsilon$-poles have cancelled, but
the infrared-sensitivity of the cross section is reflected in the
large logarithms $\ln(2 M_W/m_e)$.

\paragraph{\it Step 2: Calculation of $\hat\sigma_{LR}^{\mbox{\tiny conv}}(s)$.}
Comparing the right-hand sides of (\ref{nlophys1}) and
(\ref{nlophys2}), we obtain the radiative correction to
the conventional ``partonic'' cross section
\begin{equation}\label{eq:partonic}
\hat{\sigma}_{LR,\mbox{\tiny conv}}^{(1)}(s)
=\sigma^{(1)}_{LR}(s)-
2 \int_0^1 dx
\,\Gamma_{ee}^{\mbox{\tiny LL},(1)}(x) \,
\sigma_{LR,\mbox{\tiny Born}}(x s)\,,
\end{equation}
where $\Gamma_{ee}^{\mbox{\tiny LL},(1)}(x)$ is the $O(\alpha)$ term
in the expansion of the conventional electron structure function
provided in~\cite{Skrzypek:1992vk,Beenakker:1996kt}. In the notation
of \cite{Beenakker:1996kt} we employ the structure function with
$\beta_{exp}=\beta_s=\beta_{\mbox{\tiny H}}=\beta_e=
\frac{2 \alpha}{\pi} \left(2 \ln(\sqrt{s}/m_e)-1\right)$. To calculate
the subtraction term in (\ref{eq:partonic}) it is sufficient
to approximate $\sqrt{s}=2 M_W$ in the expression for $\beta_e$,
to set $\sigma_{LR,\mbox{\tiny Born}}(x s)$ to the
leading-order Born term (\ref{eftLOsigma}) with the replacement
of  $E$ by $E-M_W(1-x)$, and
to use $\Gamma_{ee}^{\mbox{\tiny LL},(1)}(x)$ in the limit $x\to 1$,
\begin{equation}
\Gamma_{ee}^{\mbox{\tiny LL},(1)}(x) \stackrel{x\to 1}{\to}
\frac{\beta_e}{4}\left(\,\frac{2}{[1-x]_+}+\frac{3}{2}\,\delta(1-x)
\right).
\end{equation}
 We then reintroduce the
integral over $r$, and exchange the $r$- and $x$-integration to obtain
\begin{eqnarray}
&& - 2 \int_0^1 dx
\,\Gamma_{ee}^{\mbox{\tiny LL},(1)}(x) \,
\sigma^{(0)}_{LR}(x s) =
\nonumber\\
&&\hspace*{1.5cm}-\frac{16\pi^2\alpha^2_{ew}}{27 M_W^2 s}
\,\mbox{Im} \left\{ (-i) \,\tilde\mu^{2\epsilon}
\int \frac{d^d r}{(2 \pi)^d} \frac{1}{\eta_{-} \eta_{+}}
\,\frac{\beta_e}{2}
\left[2 \ln \left(-\frac{\eta_-}{M_W}\right)+
\frac{3}{2}\right]\right\}\,,
\label{eq:collm_e}
\end{eqnarray}
which shows that $\hat\sigma_{LR}^{\mbox{\tiny conv}}(s)$ is free from
the large electron mass logarithms. 
To obtain the final form in~\eqref{eq:collm_e}
 we have shifted the integration variable $r_0$ to $E-r_0$. 
Summing (\ref{eq:totcollcross})
and (\ref{eq:collm_e}), and performing the $r$-integration,
gives the final result for the next-to-leading order radiative 
correction to the conventional ``partonic'' cross section
\begin{eqnarray}
\hat{\sigma}_{LR,\mbox{\tiny conv}}^{(1)}(s)
&=& \frac{4 \alpha^3}{27 s_w^4 s} \,\mbox{Im}
\,\Bigg\{(-1) \,\sqrt{-\frac{E+i \Gamma^{(0)}_W}{M_W}}
\,\bigg(2 \ln \bigg(-\frac{4 (E+i \Gamma^{(0)}_W)}{M_W}\bigg)
+\mbox{Re}\,\Big[c_{p,LR}^{(1,\rm fin)}\Big]
\nonumber\\
&&
+\,\frac{\pi^2}{4} +\frac{1}{2}\bigg)\Bigg\}
+\Delta \sigma^{(1)}_{\mbox{\tiny Coulomb}}
+\Delta \sigma^{(1)}_{\mbox{\tiny decay}}\,.
\label{eq:finalcross}
\end{eqnarray}

\paragraph{\it Step 3: Computation of the resummed cross section.}
The summation of collinear logarithms from initial-state radiation
is completed by performing the convolution (\ref{eq:physicalcross})
using the Born cross section and the radiative correction
(\ref{eq:finalcross}) together with the electron structure
functions from~\cite{Skrzypek:1992vk,Beenakker:1996kt}. This
constitutes our final result, which we shall discuss in detail
in the following section.

\section{NLO four-fermion production cross section}
\label{sec:results}

We now present our NLO predictions for the total cross section of
the process $e^- e^+\to \mu^-\bar\nu_\mu u \bar d \,X$ and assess
the theoretical error on the $W$-mass measurement due to the
uncertainties in the cross-section calculation.

\subsection{Input parameters and summation of $W$-width corrections}

In addition to the input
parameters~\eqref{inputs} used for the comparison of the tree
 cross section we use $\alpha_s=\alpha_s^{\overline{\rm
     MS}}(80.4\,\mbox{GeV})=0.1199$
and the masses
\begin{equation}
\begin{aligned}
m_t&=174.2\,\text{GeV},& M_H&=115\,\text{GeV},&
m_e= 0.51099892\,\text{MeV}.
\end{aligned}
\end{equation}
We use the fine structure constant $\alpha$ in the $G_\mu$
scheme everywhere including the initial-state radiation.
With these input parameters we obtain
from~\eqref{eq:Gamma1ewFS} the numerical value
of the $W$ width to NLO,
\begin{equation}
\label{eq:gamma_w}
\Gamma_W=3\,\Big(\Gamma_{W,l}^{(0)}+\Gamma_{W,l}^{(1,ew)}\Big)+
2 \,\Big(\Gamma_{W,h}^{(0)}+
\Gamma_{W,h}^{(1,ew)}\Big)
\,\delta_{\mbox{\tiny QCD}}=2.09201\,\text{GeV}.
\end{equation}
Note that we have chosen to multiply not only the leading order,
but also the electroweak correction to the hadronic decay by the
factor $\delta_{\mbox{\tiny QCD}}$ defined in
(\ref{eq:delta_qcd}). In the numerical results below we will resum
the full NLO width~\eqref{eq:gamma_w} in the effective-theory
propagator~\eqref{OmegaProp}, that is we do not perform an
expansion of the propagator in the perturbative corrections to the
matching coefficient $\Delta$. We now describe how the formula for
the NLO cross section must be modified to accomplish this
summation of the width corrections. Readers not interested in this
technical detail may move directly to the next subsection.

Leaving $\Delta =-i\Gamma_W$ unexpanded amounts
to setting $\Gamma_W^{(1)}$ to zero in the NLO tree cross
section~\eqref{eftNLOpotsigma} and to replacing $\Gamma_W^{(0)}$ by
$\Gamma_W$ wherever it appears.
Some care has then to be taken in order to obtain the correct cross section
for the flavour-specific four-fermion final state
from the calculation of the forward-scattering amplitude.
Cutting the effective-theory propagator leads to a factor
\begin{equation}
\frac{M_W \Gamma_W}{(r_0-\frac{\vec{r}^{\,2}}{2 M_W})^2+
\frac{\Gamma_W^2}{4}}\,,
\end{equation}
analogously to~\eqref{eq:im-propagator}. In the
direct calculation of the four-fermion production cross section
the numerator arises from integrating over the two-body decay phase
space, which yields the leading-order partial width. Hence, we
have to multiply all contributions to the
forward-scattering amplitude with two cut effective-theory propagators
(the potential contributions in
Section~\ref{subsec:NLOpotential}, the
Coulomb and soft radiative corrections, and the contribution from
the one-loop correction to the production operator)
by a factor $\Gamma^{(0)}_{\mu^- \bar{\nu}_\mu}\Gamma^{(0)}_{u
\bar{d}}/\Gamma_W^2 $ instead of the factor  $
\Gamma^{(0)}_{\mu^- \bar{\nu}_\mu}\Gamma^{(0)}_{u
\bar{d}}/[\Gamma_W^{(0)}]^2=1/27$ used in the tree level analysis.
In the calculation of the matching coefficient of the four-electron
production-decay operator performed in Section~\ref{subsec:N12LOhard}
the self-energy insertions on one of the two $W$ lines are treated
perturbatively, and the decay subprocess is already correctly included
at lowest order, while the other $W$ is effectively
treated in the narrow-width approximation
\begin{equation}
\frac{M_W \Gamma_W}{(k^2-M_W^2)^2+M_W^2 \Gamma_W^2}
\rightarrow \pi \frac{\Gamma_W}{\Gamma_W} \delta(k^2-M_W^2).
\end{equation}
To obtain the correct flavour-specific final state we therefore have
to include a single prefactor $\Gamma^{(0)}_{W^-\to \mu^-
\bar{\nu}}/\Gamma_W$ or $\Gamma^{(0)}_{W^+ \to u \bar{d}}/\Gamma_W$,
depending on the $W$ charge.
As shown in Table~\ref{tab:efttree}, with these
prescriptions the N$^{3/2}$LO effective-theory approximation
and the full Born cross section (in the fixed-width definition now
using (\ref{eq:gamma_w})) are  again in very good agreement,
similar to the earlier comparison, where only $\Gamma^{(0)}_W$ was
resummed in the propagator.
\begin{table}[t]
\begin{center}
\begin{tabular}{|c|c|c|c|}
\hline&
\multicolumn{2}{c}{ $\sigma(e^-e^+\to \mu^-\bar\nu_\mu u\bar d\,)$(fb)}&
\\\hline
$\sqrt{s}\,[\mbox{GeV}]$&EFT Tree (NLO) & EFT Tree (N$^{3/2}$LO)&
exact Born\\\hline
155 & 42.25 & 30.54 &33.58(1) \\\hline
158 & 65.99 & 60.83 & 61.67(2)   \\\hline
161 &154.02 &154.44 &154.19(6)  \\\hline
164 &298.6 & 303.7& 303.0(1) \\\hline
167 &400.3 &409.3  & 408.8(2)  \\ \hline
170 &469.4 & 481.7 &481.7(2)\\\hline
\end{tabular}
\end{center}
\caption{Comparison of the  numerical computation of the
  full Born result with Whizard  with successive effective-theory
  approximations as in Table~\ref{tab:efts}, but now
  the NLO decay width
  $\Gamma_W$ as given in~\eqref{eq:gamma_w} is used. }
\label{tab:efttree}
\end{table}

As already mentioned the electroweak radiative corrections are correctly
treated by multiplying the inclusive forward-scattering amplitude
by $\Gamma^{(0)}_{W^-\to \mu^- \bar{\nu}}\Gamma^{(0)}_{W^+ \to u
\bar{d}}/\Gamma_W^2 $, except for the correction to $W$ decay
itself.
These contributions are included by adding the decay correction
\begin{equation}
  \Delta\sigma^{(1)}_{\text{decay}}=
\left(\frac{\Gamma^{(1,ew)}_{\mu^-\bar\nu_\mu}}{{
\Gamma^{(0)}_{\mu^-\bar\nu_\mu} }}+
\frac{\Gamma_{u\bar d}^{(1,ew)}}{\Gamma_{u\bar d}^{(0)} }
\right) \sigma^{(0)}
\label{eq:delta-decay}
\end{equation}
instead of (\ref{eq:decaycorr}). The QCD corrections up to order
$\alpha_s^2$ are included in a similar way. Because of the large
NLO corrections to the tree cross section and the large effect of ISR,
it is sensible to apply the QCD decay correction to the full NLO
electroweak cross section. This amounts to
multiplying $\Gamma_{u\bar d}^{(0)}$,
$\Gamma_{u\bar d}^{(1,ew)}$
by the radiative correction factor $\delta_{\rm{QCD}}$ as
given in~\eqref{eq:delta_qcd}, wherever they appear, which is
consistent with the definition of the NLO $W$ width (\ref{eq:gamma_w}). 
If in addition we also account (approximately) for the
QCD decay correction to the non-resonant contributions from
Section~\ref{subsec:N12LOhard}, this is equivalent to multiplying
the entire NLO electroweak cross section by $\delta_{\rm QCD}$ and
using the QCD corrected width (\ref{eq:gamma_w}) as will be done
in the following analysis.

\subsection{NLO four-fermion production cross section in
the effective theory}

The convolution of the ``partonic'' cross section with the electron
structure functions contains integrations over partonic
center-of-mass energies far below threshold, where the
effective field theory approximation is not valid. The
EFT calculation should be matched to a full cross section calculation
below some cms energy, say $\sqrt{s}=155$ GeV, where for the full
calculation a Born treatment is sufficient, because the cross
section below threshold is small. Since the
N$^{3/2}$LO EFT approximation to the Born cross section provides
a very good approximation (except significantly below threshold),
we have found it more convenient to replace the EFT approximation
to the Born cross section convoluted according to (\ref{eq:physicalcross})
by the full ISR-improved Born cross section as generated by the Whizard
program~\cite{Kilian:2001qz} rather than to perform this matching.
To this we add the NLO radiative
correction~\eqref{eq:finalcross} (replacing the leading-order
cross section $\sigma^{(0)}$ by the full Born cross section
$\,{\sigma}_{\text{Born}}$ in the decay
correction (\ref{eq:delta-decay})),
which we also convolute with the electron
distribution functions. Here we simply cut off the integration region
$\sqrt {x_1x_2s}<155\,$GeV. The dependence on this cut-off is
negligible. Lowering it from  to $155\,$GeV to $150\,$GeV ($140$ GeV),
changes the cross section at $\sqrt s=161$ GeV
from $117.81\,$fb to $117.87\,$fb ($117.91\,$fb), while the dependence
on the cut-off for higher cms energy is even smaller.

\begin{table}[t]
\begin{center}
\begin{tabular}{|c|c|c|c|c|}
\hline&
\multicolumn{3}{c}{ $\sigma(e^-e^+\to \mu^-\bar\nu_\mu u\bar d\,X)$(fb)}&
\\\hline
$\sqrt{s}\,[\mbox{GeV}]$ & Born &
 Born(ISR)& NLO
& NLO(ISR-tree)\\\hline
158 & 61.67(2)& 45.64(2)&49.19(2)& 50.02(2)  \\
    &    &  [-26.0\%]   & [-20.2\%] &  [-18.9\%] \\\hline
161 & 154.19(6)& 108.60(4) &117.81(5) & 120.00(5)\\
    &    & [-29.6\%] &[-23.6\%]  & [-22.2\%]  \\\hline
164 & 303.0(1) & 219.7(1) &234.9(1) & 236.8(1) \\
    &     & [-27.5\%] & [-22.5\%] &[-21.8\%] \\\hline
167 & 408.8(2) & 310.2(1)&328.2(1) & 329.1(1)  \\
    &     & [-24.1\%] &   [-19.7\%]  & [-19.5\%] \\ \hline
170 & 481.7(2)& 378.4(2) &398.0(2)& 398.3(2) \\
    &      &  [-21.4\%] &  [-17.4\%]  &[-17.3\%]\\\hline
\end{tabular}
\end{center}
\caption{Two NLO implementations of the effective-theory
calculation, which differ by the treatment of initial-state
radiation compared to the ``exact'' Born cross section without (second
column) and with (third column) ISR improvement. The relative
correction in brackets is given with respect to the Born cross
section in the second column. } \label{tab:eftnlo}
\end{table}

Our result for the NLO four-fermion cross section is shown in
Table~\ref{tab:eftnlo}. The impact of radiative corrections is seen
by comparing the exact Born cross section (second column, identical
to the last column in Table~\ref{tab:efttree}), the ISR-improved
Born cross section (third column) and the NLO result
(fourth column). As is well-known initial-state radiation
results in a large negative correction (about 25\%). The size
of the genuine radiative correction is best assessed by comparing
the ``NLO'' column to the ``Born(ISR)'' column and thus seen
to be about $+8\%$. Given that we aim at a theoretical accuracy
at the sub-percent level, this is an important effect. We shall
discuss below, in Section~\ref{sec:uncertainty}, an estimate of
the remaining uncertainty of the NLO cross section.

One uncertainty is related to the fact that the conventional
implementation of ISR sums only leading logarithms, whereas a NLO
calculation of the partonic cross section should be accompanied by a
next-to-leading logarithmic resummation. Thus rather than convoluting
the full NLO partonic cross section with the structure functions as
done above and indicated in~\eqref{eq:physicalcross}, one could
equally well convolute only the Born cross section, and add the
radiative correction without ISR improvement, as done in some previous
NLO calculations~\cite{Denner:2000bj,Denner:2005es}.  Although we
favour the first option, the two implementations are formally
equivalent, because the difference is a next-to-leading logarithmic
term.  We therefore consider this difference as an estimate of the
uncertainty induced by the missing next-to-leading logarithmic
evolution of the structure functions.  To assess this uncertainty, in
the fifth column of Table~\ref{tab:eftnlo} we show the NLO
cross section based on the expression
 \begin{equation}
\label{eq:isr-tree}
\sigma_{\text{ISR-tree}}(s) = \int_0^1 dx_1 \int_0^1 dx_2 \,\Gamma_{ee}^
{\mbox{\tiny LL}}(x_1)
\Gamma_{ee}^{\mbox{\tiny LL}}(x_2)
\,\sigma_{\text{Born}}(x_1 x_2 s)+
\hat \sigma^{(1)}_{\mbox{\tiny conv}}(s),
\end{equation}
where the NLO correction to the ``partonic''
cross section, $\hat \sigma^{(1)}_{\mbox{\tiny conv}}(s)$, is 
given in~\eqref{eq:finalcross} (with $1/27$ replaced by
$\Gamma^{(0)}_{\mu^- \bar{\nu}_\mu}\Gamma^{(0)}_{u
\bar{d}}/\Gamma_W^2 $).
The comparison of the last and second-to-last columns of
Table~\ref{tab:eftnlo} shows that the difference between the two
implementations of ISR
reaches almost two percent at threshold and is therefore much larger
than the target accuracy in the per-mille range.
The difference between the two implementations becomes
smaller at higher energies and is negligible
at $\sqrt s=170$ GeV. The impact of this difference
on the accuracy of the $W$-mass measurement will be
investigated further in Section~\ref{sec:uncertainty}.

\subsection{Comparison to the full four-fermion calculation}

\begin{table}[t]
\begin{center}
\begin{tabular}{|c|c|c|c|c|}
\hline&
\multicolumn{3}{c}{
$\sigma(e^-e^+\to \mu^-\bar\nu_\mu u\bar d\,X)$(fb)}&
\\\hline
$\sqrt{s}\,[\mbox{GeV}]$ & Born & NLO(EFT) & ee4f~\cite{Denner:2005es}
& DPA~\cite{Denner:2005es} \\\hline
161& 150.05(6) &104.97(6) & 105.71(7) & 103.15(7)\\\hline
170 & 481.2(2) & 373.74(2) & 377.1(2) & 376.9(2)  \\\hline
\end{tabular}
\caption{Comparison of the strict electroweak NLO results (without
QCD corrections and ISR resummation).}
\label{tab:denner1}
\end{center}
\end{table}

We now compare the NLO prediction of the four-fermion production
process~\eqref{eq:wwprocess} obtained with the effective-theory
method to the full NLO calculation performed in~\cite{Denner:2005es} in the
complex mass scheme. For this comparison, we adjust our input
parameters to those of~\cite{Denner:2005es},
\begin{equation}
\label{eq:denner-input}
M_W=80.425\,\text{GeV} \,,\quad\Gamma_W=2.0927 \,\text{GeV}
\, ,\quad m_t=178\,\text{GeV} \,,\quad \alpha_s=0.1187,
\end{equation}
and  use $\alpha(0)=1/137.03599911$ in the relative radiative
corrections as in~\cite{Denner:2005es}. We first compare the
strict electroweak NLO calculation, i.e.~the cross section
without the QCD correction $\delta_{\text{QCD}}$ and without initial-state
radiation beyond the first-order term. In the effective-theory
calculation the corresponding radiative correction is given by
~\eqref{eq:totcollcross} omitting the
second-order Coulomb correction and the factor $\delta_{\text{QCD}}$
in the decay width. In Table~\ref{tab:denner1} the EFT result
and the result of~\cite{Denner:2005es} are shown in the columns
labelled ``NLO(EFT)'' and ``ee4f'', respectively. For comparison we
also show the results for the Born cross section and in the double-pole
approximation (``DPA'') in the implementation of~\cite{Denner:2000bj}
as quoted in~\cite{Denner:2005es}.
The main observation is that the difference between the EFT and
the full four-fermion calculation is only  $0.7\%$ at $\sqrt s=161$
GeV and grows to  about $1\%$ at $\sqrt s=170$ GeV.

Next, in Table~\ref{tab:denner2}, we compare to the
full result including the QCD correction and the resummation of
ISR corrections with~\cite{Denner:2005es}. Here we implement
the QCD correction as in~\cite{Denner:2005es} by multiplying
the entire electroweak NLO result by the overall
factor $(1+\alpha_s/\pi)$. Furthermore, we
include ISR corrections only to the Born cross section
as in~\eqref{eq:isr-tree}, in agreement with the treatment
of~\cite{Denner:2005es}. Again the second-order Coulomb correction
is set to zero, because \cite{Denner:2005es} does not include any
two-loop effects. As before, the Table shows the two NLO
calculations, the Born cross section (now ISR improved)
and the double-pole approximation.
The discrepancy between the EFT calculation and the full four-fermion
calculation is around $0.6\%$ at threshold. The EFT approximation
is significantly better than the double-pole approximation directly
at threshold, while at higher energies the quality of the DPA
improves relative to the EFT approximation, since no threshold
expansion is performed in the DPA.

\begin{table}[t]
\begin{center}
\begin{tabular}{|c|c|c|c|c|}
\hline&
\multicolumn{3}{c}{
$\sigma(e^-e^+\to \mu^-\bar\nu_\mu u\bar d\,X)$(fb)}&
\\\hline
$\sqrt{s}\,[\mbox{GeV}]$& Born(ISR) & NLO(EFT) &
ee4f~\cite{Denner:2005es} & DPA~\cite{Denner:2005es} \\\hline
161 &  107.06(4) &117.38(4)   & 118.12(8)& 115.48(7)\\\hline
170 &  381.0(2) & 399.9(2)   & 401.8(2) & 402.1(2)  \\\hline
\end{tabular}
\caption{Comparison of NLO results with
QCD corrections and ISR resummation included.}
\label{tab:denner2}
\end{center}
\end{table}

\subsection{\boldmath Theoretical error of the $M_W$ determination}
\label{sec:uncertainty}

The $W$ mass will probably be determined by measuring the
four-fermion production cross section at a few selected
cms energies near the $W$ pair-production threshold.
In this section we estimate the error on the
$W$ mass from various sources of theoretical uncertainty.
To this end we assume that measurements $O_i$ will be taken at
$\sqrt{s}=160,161,162,163,164\, \mbox{GeV}$, and at
$\sqrt{s} = 170 \, \mbox{GeV}$, and that the measured values
coincide with our NLO calculation (labelled ``NLO(EFT)'' in
Table~\ref{tab:eftnlo}) corresponding to the
$W$ pole mass $M_W=80.377\, \mbox{GeV}$. We denote
by $E_i(\delta M_W)$ the cross section values at the six
cms energy points for any other theoretical calculation
of four-fermion production as a function of the
input $W$ mass $80.377\, \mbox{GeV}+\delta M_W$, and
determine the minimum of
\begin{equation}
\chi^2(\delta M_W)=\sum_{i=1}^6
\frac{\left(O_i-E_i(\delta M_W)\right)^2}{2 \sigma_i^2}\,.
\end{equation}
For simplicity we assume that each point carries the same
weight, so $\sigma_i\equiv \sigma$ is an arbitrary constant of
mass dimension $-2$. (We checked that a 
more realistic assignment $\sigma_i\sim \sqrt{O_i}$
does not lead to significantly different results.)
The value of $\delta M_W$ at which
$\chi^2(\delta M_W)$ attains its minimum provides an estimate
of the difference in the measured value of $M_W$ due to the
different theoretical cross section inputs, $O_i$ and $E_i$.
For instance if $E_i(\delta M_W)$ is the ISR-improved
Born cross section (labelled ``Born(ISR)'' in Table~\ref{tab:eftnlo}),
we obtain $\delta M_W = -201\,$MeV, which tells
us that comparing measurements to a theoretical calculation
without the genuine radiative corrections would result in a value
of $M_W$ which is about 200 MeV too low. The NLO calculation
is therefore crucial for an accurate $M_W$ determination.
Next we attempt to estimate whether it is accurate enough.

\paragraph{\it Treatment of initial-state radiation.}
A look at the last two columns of Table \ref{tab:eftnlo} reveals
that two different implementation of ISR, which are formally
equivalent at the leading-logarithmic level, can lead
to differences in the predicted cross section of $2 \%$ at
$\sqrt{s}=161\,$GeV, where the sensitivity to $M_W$ is largest. 
We take this as a measure for the uncertainty caused by the missing
next-to-leading logarithmic corrections to the structure function.
To
estimate the error on $M_W$ caused by this uncertainty, we apply
the procedure discussed above and find
\begin{equation}
[\delta M_W]_{\rm ISR} \approx 31\, \mbox{MeV}.
\end{equation}
This large error could be avoided by measuring the cross
section predominantly around $170\,$GeV rather than around
$162\,$GeV, but the sensitivity to $M_W$ is significantly smaller
at higher energies (see Figure~\ref{fig:massuncertainty} below).
Thus, this error should be eliminated by a consistent treatment
of the electron structure functions at the next-to-leading logarithmic
level, in which all NLL corrections are taken into account by convoluting the 
NLO cross section with the NLL structure functions.
 A related effect concerns the choice of scheme and scale
of the electromagnetic coupling. The difference in the cross section
between using $\alpha(0)$ and $\alpha$ in the $G_\mu$-scheme in the
radiative correction (including, in particular, initial-state radiation)
is about $1\%$, which translates into another error of about
$15\,$MeV in the $W$ mass. 
 The scale ambiguity
of the coupling used in initial-state radiation can be resolved only in
the context of a next-to-leading logarithmic resummation which takes the 
evolution of $\alpha$ between $m_e$ and $\Gamma_W$ into account. 
On the other hand,  the typical scales in the short-distance
cross section are at least $\Gamma_W \approx 2\,$GeV, so 
that $\alpha$ in the $G_\mu$ scheme is more appropriate
than the low-energy electromagnetic coupling in the radiative
correction to the short-distance cross section, since it is numerically 
close to the running coupling at  $2\,$GeV.

\paragraph{\it Uncalculated corrections to the ``partonic''  cross section.}
The leading missing higher-order terms in the expansion in
$\alpha$ and $\delta$ are $\mbox{N}^{3/2}$LO corrections to the
forward-scattering amplitude from four-loop potential
diagrams (third Coulomb correction), three-loop diagrams with two
potential loops and one soft loop (interference of single-Coulomb and
soft radiative corrections), two-loop potential diagrams with
$O(\alpha)$ matching coefficients or $O(\delta)$
higher-dimensional production operators, and the  $O(\alpha)$
correction to the matching coefficients of the four-electron
production-decay operators. The latter is expected to be the largest
of these contributions, in particular since the non-resonant
N$^{1/2}$LO contributions are large at the Born level ($\sim 40\%$ at
threshold, see Table~\ref{tab:efts}). Presumably, this contribution is
also the origin of the $0.6 \%$ difference between the EFT result
``NLO(EFT)'' and the full four-fermion calculation ``ee4f''
at $\sqrt{s}=161\,$GeV in Table~\ref{tab:denner2}.
A rough estimate of this correction to the helicity-averaged
cross section is
\begin{equation}\label{eq:N3half4ferm}
\Delta \hat\sigma = \frac{\alpha^4}{27 s_w^8 s} \,\mathcal{K},
\end{equation}
where $\mathcal{K}$ is an $s$-independent constant of order
1. (In fact, if we attributed the difference between our calculation
(``NLO(EFT)'') and that of \cite{Denner:2005es} (``ee4f'')
at $\sqrt{s}=161\,$GeV exclusively to this contribution, we
would obtain $\mathcal{K}=0.96$.) Thus, we choose
$\mathcal{K}=1$, add (\ref{eq:N3half4ferm}) to the
``NLO(EFT)'' calculation, and minimize the $\chi^2$ function.
From this we obtain an error
\begin{equation}
[\delta M_W]_{\rm non-res} \approx 8\, \mbox{MeV}. \label{error2}
\end{equation}
The second largest uncalculated correction to the partonic
cross section is expected to come from diagrams with single-Coulomb
exchange and a soft photon or a hard correction to the production
vertex. A naive estimate of the sum of the two terms is
\begin{equation} \label{eq:N3halfcoulomb}
\Delta \hat \sigma =
\frac{\hat \sigma^{(1)}_{LR}-\Delta\sigma^{(1)}_{\mbox{\tiny Coulomb}}-
\Delta\sigma^{(1)}_{\mbox{\tiny decay}}}{\sigma_{LR}^{(0)}}
\,\Delta\sigma^{(1)}_{\mbox{\tiny Coulomb}}\,,
\end{equation}
where the quantities involved have been defined in Section
\ref{sec:radcor}. Estimating the
corresponding uncertainty on the $W$ mass as before, we find
\begin{equation}
[\delta M_W]_{\rm Coulomb \times (hard+soft)} \approx - 5\,
\mbox{MeV}. \label{error3}
\end{equation}
Adding the two errors we conclude that the uncertainty on
$M_W$ due to uncalculated higher-order effects in the effective
field theory method is about $10-15\,$MeV. Thus, to reach a
total error of $\sim 6\, \mbox{MeV}$ requires the inclusion of
at least some $\mbox{N}^{3/2}$LO corrections in the EFT approach.
The larger of the two errors estimated above,
due to the electroweak correction to
production-decay operator, can be removed by using the full
NLO four-fermion calculation, where this correction is included.
\begin{figure}[t]
\begin{center}
\vspace{-1cm}
\includegraphics[width=0.7\linewidth]{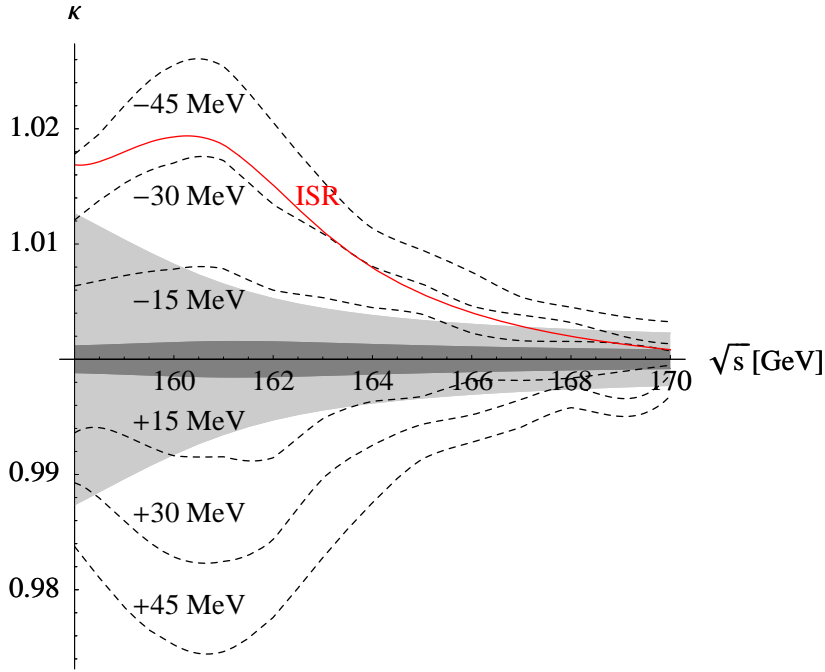}
\vspace{-1cm}
\end{center}
\caption{$W$-mass dependence of the total cross section. All the
cross sections are normalized to $\sigma(s,M_W=80.377\,\mbox{GeV})$.
See text for explanations.}
\label{fig:massuncertainty}
\end{figure}

\paragraph{\it Summary.} The discussion above is summarized in
Figure~\ref{fig:massuncertainty}, where we plot
$\kappa = \sigma(s,M_W+\delta M_W)/\sigma(s,M_W)$ for
different values of $\delta M_W$ as function of the cms energy,
$\sigma$ being our NLO result, ``NLO(EFT)''. The relative change in
the cross section is shown as dashed lines for $\delta M_W=\pm
15,\pm 30, \pm 45 \, \mbox{MeV}$. The shape of these curves shows
that the sensitivity of the cross section to the $W$ mass is
largest around the nominal threshold $\sqrt{s}\approx 161\,$GeV,
as expected, and rapidly decreases for larger $\sqrt{s}$. (The loss
in sensitivity is partially compensated by a larger cross section,
implying smaller statistical errors of the anticipated experimental
data.)

The dark-shaded area in Figure~\ref{fig:massuncertainty} corresponds
to the uncertainty on the cross section from
(\ref{eq:N3halfcoulomb}), while the light-shaded area adds
(linearly) the uncertainty from (\ref{eq:N3half4ferm}). The
theoretical error decreases with $\sqrt{s}$, since $\Delta\sigma$
in (\ref{eq:N3half4ferm}) is roughly energy-independent, while
$\sigma$ increases. The largest current uncertainty is, however,
due to ambiguities in the implementation of ISR. The solid (red)
curve gives the ratio of the two different implementations of
ISR, NLO(EFT) vs. NLO(ISR-tree), both evaluated at $M_W=80.377\,
\mbox{GeV}$. As mentioned above, we do not consider this
as a fundamental problem, since this uncertainty can be removed
with further work on a next-to-leading-logarithmic ISR resummation
that will be required for many other processes at a high-energy
$e^- e^+$ collider as well.

\section{Conclusion}
\label{sec:conclude}

We performed a dedicated study of four-fermion production near the
$W$ pair-production threshold in view of the importance of this
process for an accurate determination of the $W$-boson mass. Our
theoretical study of radiative and finite-width corrections was
motivated by a corresponding experimental
study~\cite{Wilson:2001aw} which showed that the planned
high-luminosity linear collider might allow a measurement of $M_W$
with an error of only $6\,$MeV from the threshold region. Our
calculation, and the good agreement with the full NLO four-fermion
cross section calculation of~\cite{Denner:2005es}, demonstrates that 
accurate theoretical calculations are feasible and available in the
threshold region. With regard to the mass determination, we find:
\begin{itemize}
\item A resummation of next-to-leading collinear logarithms from
initial-state radiation is mandatory to reduce the error on
$M_W$ below the 30 MeV level.
\item The NLO partonic cross-section calculation in the effective
theory approach implies a residual error of about 10~--~15 MeV
on $M_W$. The largest missing N$^{3/2}$LO effect is probably due to
the electroweak correction to the (non-resonant) production-decay
vertex, which is included in the full NLO four-fermion calculation,
and can thus be eliminated.
\end{itemize}
It is forseeable that both items can be removed, so we conclude that there is
no fundamental difficulty in reducing the theoretical error in
the $W$ mass determination from the threshold region to about
5 MeV.

The calculation presented here is also the first NLO calculation of a
realistic process in unstable-particle effective theory, since
\cite{Beneke:2003xh,Beneke:2004km} discussed the case of a single
resonance in a gauged Yukawa model. Comparison of our results for
four-fermion production with numerical integrations of the Born matrix
elements and the radiative correction shows good convergence of the
EFT expansion, and very good agreement once the first subleading term
in each essential region (potential/resonant, hard/non-resonant) is
included. The EFT approach provides a consistent treatment of
finite-width effects that can in principle be extended systematically
to higher orders.  Our final results take the form of compact analytic
formulae, which has to be compared to the numerical and technical
challenges~\cite{Denner:2005es} of the full NLO four-fermion cross
section calculation. However, it should be mentioned that our
calculation is restricted to the inclusive cross section, while a more
flexible treatment of the final-state phase space is obviously
desirable. This requires either applying effective-theory methods to
four-fermion production amplitudes rather than the forward-scattering
amplitude, or the consideration of specific cuts such as corresponding
to invariant-mass distributions that allow for a semi-inclusive
treatment. Interesting developments in this direction have recently
been reported for top-quark pair production \cite{Fleming:2007qr}.

\vspace*{0.5em}
\noindent
\subsubsection*{Acknowledgement}
This work is supported in part
by the DFG Sonder\-forschungsbereich/Transregio~9
``Computergest\"utzte Theoretische Teilchenphysik'', the
DFG Graduiertenkolleg ``Elementar\-teil\-chen\-physik an der
TeV-Skala'', and the European Community's Marie-Curie
Research Training Network under contract MRTN-CT-2006-035505 `Tools
and Precision Calculations for Physics Discoveries at Colliders'.

\appendix

\section{Coefficients of non-resonant contributions}
\label{ap:4ferm}

In this appendix we list the explicit expressions of the
remaining coefficients $C^f_{i,h}$ and $K^f_i$ in (\ref{eq:treehard}).
The functions $C^f_{i,h}$ are known analytically, and
contain all the $s$-dependence of the cross section (except for
the overall factor $1/s$). They are determined by the
photon and $Z$ propagators and electroweak couplings.
In the limit of vanishing fermion masses the only
helicity configurations contributing to the cross section are $h=LR\,,RL$.
The coefficients $K^f_i$ are $s$-independent, and result from
dimensionally regularized cut loop integrals. Typically the last integration
is performed numerically, after the subtraction of the singular terms which are
integrated analytically, though some analytic
results can be obtained. The results given below contain the
contribution of the diagrams h4-h7 in Figure~\ref{fig:hardcuts} including
their complex conjugates, except for cut h6, where the complex conjugate
is the diagram itself, and cut h7, where the symmetric diagram
is automatically taken into account by summing over the four
flavours.

Only the configuration $e^-_L e^+_R$ contributes to the cut diagram h4:
\begin{eqnarray}
&&C^{f}_{h4,LR}=3 M_W^2 s_w^2 \left(-\frac{Q_f}{s}+
\frac{C_e^L C_f^L}{s-M_Z^2}\right)\,,\nonumber\\
&&K^{u}_{h4}= K^{\nu_\mu}_{h4}=-0.266477\,,
\hspace{0.5 cm} K^{d}_{h4}=K^{\mu}_{h4}=0.190394\,,
\end{eqnarray}
where $Q_f$ and $C^L_f=\frac{I_{W,f}^3-s_w^2 Q_f}{s_w c_w}$ are
the couplings of left-handed fermions to $\gamma$ and $Z$. $Q_f$
always denotes the charge of the particle (not the anti-particle) in units of
the positron charge.
For the cut diagram h5 we have
\begin{eqnarray}
&&C^{f}_{h5,h}=9 M_W^4 s_w^4 \left(-\frac{Q_f}{s^2}
+\frac{C^{h}_e C^L_f}{s (s-M_Z^2)}
+\frac{c_w}{s_w}\frac{Q_f C_e^{h}}{s (s-M_Z^2)}
-\frac{c_w}{s_w}\frac{{C^{h}_e}^2 C^L_f}{(s-M_Z^2)^2}\right)\,,\nonumber\\
&&K^{u}_{h5}=K^{\nu_\mu}_{h5}=0.455244\,,\hspace{0.5 cm}
  K^{d}_{h5}=K^{\mu}_{h5}= -0.455244\,,
\end{eqnarray}
where $C^{LR}_e=C^L_e$ and $C^{RL}_e=C^R_e=-\frac{s_w}{c_w} Q_e$.
In this case both left-handed and right-handed incoming fermions
contribute ($h=LR,RL$), but only left-handed internal fermions.
The coefficients of h6 are
\begin{eqnarray}
&&C^{f}_{h6,h}=9 M_W^4 s_w^4 \left(-\frac{Q_f}{s}
+\frac{C_e^{h} C^L_f}{s-M_Z^2}\right)^2\,,\nonumber\\
&&K^{u}_{h6} = K^{d}_{h6}=K^{\mu}_{h6}= K^{\nu_\mu}_{h6}= 0.0804075\,,
\end{eqnarray}
while for h7 we get
\begin{eqnarray}
&&C^{f}_{h7,h}=9 M_W^4 s_w^4 \left(\frac{ Q_f \bar{Q}_f}{s^2}
-\frac{Q_f C^{h}_e \bar{C}^L_f}{s (s-M_Z^2)}-
\frac{\bar{Q}_f C_e^{h} C^L_f}{s (s-M_Z^2)}
+\frac{{C^{h}_e}^2 C^L_f \bar{C}^L_f}{(s-M_Z^2)^2}\right)\,,
\nonumber\\
&&K^{u}_{h7} = K^{d}_{h7} = K^{\mu}_{h7} = K^{\nu_\mu}_{h7} = 0.0213082\,,
\end{eqnarray}
where $Q_f,\, \bar{Q}_f$ and $C^L_f,\, \bar{C}^L_f$ are the
couplings to $\gamma$ and $Z$ of the particles in the
same SU(2) doublet (i.e. $\mu,\, \nu_{\mu}$ and $u,\,d$).

\section{Hard one-loop coefficients}
\label{ap:hard1loop}

\def\LL{\ell}

We give here the explicit analytic results for the hard one-loop
coefficients appearing in Section~\ref{subsec:hard}.

\subsection{Production vertices}
\label{app:production}
The general $e^- e^+ \to W^-W^+$ production operator we are concerned
with in this appendix reads
\begin{equation}
  {\cal O}_p = \frac{\pi\alpha_{ew}}{ M_W^2} C_{p}
    \left(\bar{e} \gamma^{[i} n^{j]} e \right)
    \left(\Omega_-^{\dagger i} \Omega_+^{\dagger j}\right)\,,
    \label{Pgen}
\end{equation}
where $C_{p} = C_{p,h}$ is the hard matching coefficient and $h=LR,RL$
refers to the helicity of the incoming leptons ($e^-_L e^+_R$ or
$e^-_R e^+_L$). Starting with $e^-_L e^+_R \to W^-W^+$, the matching
coefficient at tree level is equal to $1$, as can be read off
(\ref{LPlead}). At NLO we have
\begin{equation}
C_{p, LR} = 1+ C_{p, LR}^{(1)} +\cO{\alpha^2} \equiv
 1+ \frac{\alpha}{2\pi} c_{p, LR}^{(1)}+\cO{\alpha^2}\,,
\end{equation}
where $C_{p, LR}^{(1)}$ is the coefficient in (\ref{LPNLO}).
Before renormalization the NLO short-distance coefficient reads
\begin{eqnarray}
  \label{eq:PV0LR}
c_{p, LR}^{(1),\, {\rm bare}} &=&
-\frac{1}{\epsilon ^2}
    \left(-\frac{4M_W^2}{\mu^2}\right)^{\!-\epsilon}
+\frac{8 c_w^4+10 c_w^2+1}{8 c_w^2 s_w^2\ \epsilon }
    \left(-\frac{4M_W^2}{\mu^2}\right)^{\!-\epsilon}
\nonumber \\
&+&\frac{\left(2 c_w^2-1\right) \left(24 c_w^4+16 c_w^2-1\right) \
   M_W^2\ C_0\left(0,M_W^2,-M_W^2,0,M_Z^2,M_W^2\right)}
   {8 c_w^4 s_w^4} \nonumber \\
&-&\frac{\left(2 c_w^2-1\right) \
   M_W^2\ C_0\left(0,4 M_W^2,0,0,M_Z^2,M_Z^2\right)}
   {2 c_w^4 s_w^2}\nonumber \\
&-&\frac{\left(\left(c_w^4+17 c_w^2-16\right) M_H^2+M_W^2\right) \
   M_W^2\ C_0\left(-M_W^2,M_W^2,0,0,0,M_W^2\right)}
   {4 M_H^2 \ s_w^2}\nonumber \\
&+&\frac{\left(M_H^2+M_W^2\right) \
   M_W^2\ C_0\left(-M_W^2,M_W^2,0,0,M_H^2,M_W^2\right)}
   {4 M_H^2\ s_w^2}\nonumber \\
&-&\frac{\left(2 c_w^8+32 c_w^6+32 c_w^4-11 c_w^2-16\right) \
   M_W^2\ C_0\left(-M_W^2,M_W^2,0,0,M_Z^2,M_W^2\right)}
   {8 c_w^2 s_w^4}\nonumber \\
&+&\frac{3 \left(33-46 c_w^2\right) \
   M_W^2\ C_0\left(M_W^2,-M_W^2,0,0,0,M_W^2\right)}
   {8 s_w^4}\nonumber \\
&+&\frac{\left(4 c_w^4-1\right) \left(14 c_w^6+15 c_w^4-2 c_w^2-1\right) \
   M_W^2\ C_0\left(M_W^2,-M_W^2,0,0,0,M_Z^2\right)}
   {16 c_w^8 s_w^4}\nonumber \\
&-&\frac{\left(1-2 c_w^2\right)^2 \left(c_w^2+1\right) \
   \left(4 c_w^2+1\right)^2
    M_W^2\ C_0\left(4 M_W^2,0,0,0,0,M_Z^2\right)}
    {16 c_w^8 s_w^2}\nonumber \\
&-&\frac{25 M_W^2\ C_0\left(4 M_W^2,0,0,0,0,M_W^2\right)}{4 s_w^2}
   + \frac{M_W^2\ \LL\left(M_W^2,M_W^2,M_H^2\right)}{4 M_H^2\ s_w^2}
   \nonumber \\
&+&\frac{\left(-168 c_w^8-214 c_w^6+56 c_w^4+32 c_w^2-3\right) \
   \LL\left(M_W^2,M_W^2,M_Z^2\right)}
   {24 c_w^2 \left(1-4 c_w^2\right) s_w^2}\nonumber \\
&+&\frac{\left(1-2 c_w^2\right) \left(8 c_w^4+c_w^2+3\right)
   \LL\left(4 M_W^2,M_Z^2,M_Z^2\right)}{6 c_w^2 s_w^2}\nonumber \\
&+&\frac{3 \left(c_w^2+1\right)
   \ln\left(\frac{M_W^2}{M_Z^2}+1\right)}{16 \ c_w^6}
+ \frac{\left(1-2 c_w^2\right) \left(64 c_w^4+4 c_w^2+1\right)
   \ln \left(\frac{4 M_W^2}{M_Z^2}-1\right)}{24 c_w^4}\nonumber \\
&+&\frac{\left(
    -512 c_w^{10}+1536 c_w^8-672 c_w^6+44 c_w^4+3 c_w^2-3\right)
    \ln\left(\frac{M_Z^2}{M_W^2}\right)}
    {48 c_w^4 \left(1-4 c_w^2\right) s_w^2}\nonumber \\
&+&\frac{\left(
    -128 c_w^{10}+304 c_w^8+144 c_w^6-38 c_w^4+9 c_w^2+3\right)\ln 2}
    {24 c_w^6 s_w^2}\nonumber \\
&+&\frac{96 c_w^6 - \left( 10 - 2 s_w^2\pi ^2\right) c_w^4
         -9 c_w^2-6}{24 c_w^4 s_w^2}
\nonumber \\
&-& \frac{\left(128 c_w^8-64 c_w^6+4 c_w^4+23 c_w^2+5\right)\, i \pi}
   {48 c_w^4 s_w^2}
\,,
\end{eqnarray}
where all functions appearing in the above expression,
$C_0(p_1^2,p_2^2,p_3^2,m_1^2,m_2^2,m_3^2)$ and $\LL(q^2,M_1^2,M_2^2)$,
are known analytically and are supplied in
Appendix~\ref{app:integrals}.
The counterterms in the $G_\mu$ scheme are computed from
(\ref{eq:ctterm}) and are given by
\begin{eqnarray}
  \label{eq:CTLR}
c_{p, LR}^{(1),\, {\rm ct}} &=&
\frac{4 c_w^4-22 c_w^2-1}{8 c_w^2 s_w^2\ \epsilon }
  \left(-\frac{4M_W^2}{\mu^2}\right)^{-\epsilon}
- \frac{\left(M_H^4-3 M_W^2 M_H^2+6 M_W^4\right) \
  \LL\left(M_W^2,M_H^2,M_W^2\right)}{12 M_W^4 \ s_w^2} \nonumber \\
&-&\frac{\left(M_H^2-5 M_W^2\right) \
  \LL\left(0,M_H^2,M_W^2\right)}{12 M_W^2 \ s_w^2}
 -\frac{\left(8 c_w^4+27 c_w^2-5\right) \
  \LL\left(0,M_W^2,M_Z^2\right)}{12 c_w^2 s_w^2}\nonumber \\
&+& \frac{\left(42 c_w^4-11 c_w^2-1\right) \
  \LL\left(M_W^2,M_W^2,M_Z^2\right)}{12 c_w^4 s_w^2}\nonumber \\
&-&\frac{\left(M_H^4-4 M_W^2 M_H^2+12 M_W^4\right) \
   \DB0\left(M_W^2,M_W^2,M_H^2\right)}{24 M_W^2 s_w^2}\nonumber \\
&+&\frac{\left(48 c_w^6+68 c_w^4-16 c_w^2-1\right) M_W^2 \
  \DB0\left(M_W^2,M_W^2,M_Z^2\right)}{24 c_w^4 s_w^2}\nonumber \\
&+& \frac{\left(2 M_H^4-3 M_H^2 M_W^2+2 M_W^4\right)
 \ln\left(\frac{M_H^2}{M_W^2}\right)}{24 M_W^2 (M_H^2-M_W^2) s_w^2}
 + \frac{M_H^4}{12 M_W^4\ s_w^2}
 -\frac{3 M_H^2}{16 M_W^2\ s_w^2} \nonumber \\
&-&\frac{3 m_t^2\left(m_t^4-M_W^4\right)
  \ln\left(1-\frac{M_W^2}{m_t^2}\right)}{4 M_W^6\ s_w^2}
 -\frac{3 m_t^2}{8 M_W^2 \ s_w^2}
 -\frac{3 m_t^4}{4 M_W^4 s_w^2} \nonumber \\
&-&\frac{\left(12 c_w^8-72 c_w^6+26 c_w^4-15 c_w^2-2\right)
  \ln\left(\frac{M_Z^2}{M_W^2}\right)}{24 c_w^4 s_w^4}
 + \frac{\left(4 c_w^4-22 c_w^2-1\right)\ln 2}{4 c_w^2 s_w^2}
\nonumber \\
&+&\frac{2 (35-6 i \pi ) c_w^6+(-112+66 i \pi ) c_w^4
   +(13+3 i \pi ) c_w^2+2}{24 c_w^4 s_w^2}
\,.
\end{eqnarray}
The full renormalized coefficient is obtained by adding bare result
and counterterms
\begin{equation}
c_{p, LR}^{(1)} =
c_{p, LR}^{(1),\, {\rm bare}} + c_{p, LR}^{(1),\, {\rm ct}} .
\end{equation}
The poles of $c_{p, LR}^{(1)}$ are given explicitly in
(\ref{eq:CLR}) and cancel once one takes into account soft and
initial-state collinear radiation.

Turning to the $e^-_R e^+_L \to W^-W^+$ case, the matching coefficient
$C_{p, RL}$ vanishes at tree level, as can be seen from
(\ref{LPlead}).  The NLO correction is therefore finite. We have
\begin{equation}
C_{p, RL} = C_{p, RL}^{(1)}+\cO{\alpha^2} =
\frac{\alpha}{2\pi} c_{p, RL}^{(1)}+\cO{\alpha^2}
\,,
\end{equation}
where $C_{p, RL}^{(1)}$ is the coefficient in (\ref{LPNLO}).
We find
\begin{eqnarray}
  \label{eq:PV0RL}
c_{p, RL}^{(1)} &=&
\frac{4 s_w^2\ M_W^2\ C_0\left(0,M_W^2,-M_W^2,0,M_Z^2,M_W^2\right)}
   {c_w^2 \left(2 c_w^2-1\right)}
- \frac{2  s_w^2\ M_W^2\ C_0\left(0,4 M_W^2,0,0,M_Z^2,M_Z^2\right)}
  {c_w^4 \left(2 c_w^2-1\right)}\nonumber \\
&+&\frac{\left(24 c_w^4+20 c_w^2-5\right) s_w^2\
   \LL\left(M_W^2,M_W^2,M_Z^2\right)}
     {3 c_w^2 \left(2 c_w^2-1\right) \left(4 c_w^2-1\right)}
-\frac{2 \left(8 c_w^4+c_w^2+3\right) s_w^2\
    \LL\left(4 M_W^2,M_Z^2,M_Z^2\right)}
     {3 c_w^2\left(2 c_w^2-1\right)}\nonumber \\
&+&\frac{\left(64 c_w^4+4c_w^2+1\right)  s_w^2\
   \ln\left(\frac{4 M_W^2}{M_Z^2}-1\right)}{12c_w^4}
 +\frac{\left(64 c_w^6-48 c_w^4-24 c_w^2+5\right)  s_w^2\
   \ln\left(\frac{M_Z^2}{M_W^2}\right)}
   {3 c_w^2 \left(2 c_w^2-1\right) \left(4 c_w^2-1\right)}\nonumber \\
&-&\frac{16 s_w^2 \ln 2}{3}
- \frac{\left(32 c_w^4+4 c_w^2+1\right)s_w^2 \ i \pi}{12 c_w^4}\>.
\end{eqnarray}

\subsection{Virtual corrections to $W$ decay}
\label{app:decay}

The decay of
a $W$ boson is implemented in the effective theory
analogous to the production~\cite{Beneke:2004xd}. There are decay
operators with collinear fields describing the decay products of the
non-relativistic vector boson. For the flavour-specific decays under
consideration we have up to NLO
\begin{equation}
\label{eq:decayop}
{\cal O}_d = - \frac{g_{ew}}{2\sqrt{M_W}} \left( C_{d, l}\,
\Omega^i_{-} \bar{\mu}_{c_3,L}
\gamma^i \nu_{c_4,L}  + C_{d, h}\,  \Omega^i_{+} \bar{u}_{c_3,L}
\gamma^i d_{c_4,L} \right) .
\end{equation}
These operators would be needed for the calculation of the
$e^- e^+ \to \mu^-\, \bar{\nu}_\mu\, u\, \bar{d}$
scattering {\em amplitude}  in the effective
theory. However, for the total cross section (or the forward
scattering amplitude) the directions $c_3$, $c_4$ of the decay products
will be integrated over and, as indicated in (\ref{eq:master}), there
is no need to introduce collinear fields $\bar{\mu}_{c_3,L}$,
$\nu_{c_4,L}$, $\bar{u}_{c_3,L}$ and $d_{c_4,L}$ in the effective
theory. The matching coefficients of the decay operators enter only
indirectly through $\Delta^{(2)}$.
The virtual correction to the
$W$ decay width is related to the coefficient functions
of the decay operators. Ignoring QCD corrections, at NLO we have
\begin{eqnarray}
C_{d,l} &=& 1 + C_{d,l}^{(1)} +\cO{\alpha^2}
         \equiv 1 + \frac{\alpha}{2\pi}
c_{d,l}^{(1)} +\cO{\alpha^2}\,, \nonumber \\
C_{d,h} &=& 1 + C_{d,h}^{(1)} +\cO{\alpha^2}
         \equiv 1 + \frac{\alpha}{2\pi}
c_{d,h}^{(1)} +\cO{\alpha^2} \,.
\end{eqnarray}
We give here the explicit results for the electroweak
corrections. The unrenormalized one-loop correction to the leptonic
decay vertex reads
\begin{eqnarray}
  \label{eq:LVD0}
c_{d, l}^{(1),\, {\rm bare}} &=&
-\frac{1}{2 \epsilon ^2}
  \left(\frac{M_W^2}{\mu^2}\right)^{\!-\epsilon}
+ \frac{8 c_w^4+2 c_w^2+1}{8 c_w^2 s_w^2\ \epsilon }
 \left(\frac{M_W^2}{\mu^2}\right)^{\!-\epsilon} \nonumber \\
&+&\frac{\left(c_w^2+1\right)^2 \left(2 c_w^2-1\right) \ M_W^2 \
  C_0\left(M_W^2,0,0,0,0,M_Z^2\right)}{4 c_w^6 s_w^2}\nonumber \\
&+&\frac{\left(c_w^2+2\right) \ M_W^2\
   C_0\left(M_W^2,0,0,M_W^2,M_Z^2,0\right)}{s_w^2}\nonumber \\
&+&\frac{\left(2 c_w^2+1\right) \
  \LL\left(M_W^2,M_W^2,M_Z^2\right)}{2 s_w^2}
- \frac{\left(4 c_w^6-2 c_w^4+1\right)
  \ln\left(\frac{M_Z^2}{M_W^2}\right)}{4 c_w^4 s_w^2}\nonumber \\
&-&\frac{-\left(24+\pi ^2\right) c_w^6+ (\pi^2 -18\, i \pi ) \
   c_w^4-3 i \pi  c_w^2+6 i \pi +6}{24 c_w^4 s_w^2} \>,
\end{eqnarray}
and the corresponding counterterms computed from (\ref{eq:ctterm})
are
\begin{eqnarray}
  \label{eq:LVDCT}
c_{d, l}^{(1),\, {\rm ct}} &=&
\frac{c_{p, LR}^{(1),\, {\rm ct}}}{2}
-\frac{2 c_w^2+1}{16 c_w^2 s_w^2\ \epsilon}
\left(\frac{M_W^2}{\mu^2}\right)^{-\epsilon}
+\frac{\ln \left(\frac{M_Z^2}{M_W^2}\right)}{16 c_w^2 s_w^2} \
+\frac{2 c_w^2+1}{32 c_w^2 s_w^2} \>.
\end{eqnarray}
Similarly the NLO bare correction to the hadronic vertex is given by
\begin{eqnarray}
  \label{eq:HDV0}
c_{d, h}^{(1),\, {\rm bare}} &=&
-\frac{1}{2\epsilon^2}
      \left(\frac{M_W^2}{\mu^2}\right)^{-\epsilon}
+\frac{2}{9\epsilon^2}
      \left(-\frac{M_W^2}{\mu^2}\right)^{-\epsilon}\nonumber \\
&+&\frac{(1+2 c_w^2)(1+32 c_w^2)}{72 s_w^2 c_w^2\ \epsilon}
  \left(\frac{M_W^2}{\mu^2}\right)^{-\epsilon}
+ \frac{1}{3\epsilon}
      \left(-\frac{M_W^2}{\mu^2}\right)^{-\epsilon} \nonumber \\
&+&\frac{\left(8 c_w^8+18 c_w^6+11 c_w^4-1\right) \ M_W^2\
  C_0\left(M_W^2,0,0,0,0,M_Z^2\right)}{36 c_w^6 s_w^2}\nonumber \\
&+&\frac{\left(c_w^2+2\right) \ M_W^2\
C_0\left(M_W^2,0,0,M_W^2,M_Z^2,0\right)}{s_w^2}\nonumber \\
&+&\frac{\left(2 c_w^2+1\right) \
   \LL\left(M_W^2,M_W^2,M_Z^2\right)}{2 s_w^2}
-  \frac{\left(20 c_w^6+6 c_w^4+1\right)
   \ln\left(\frac{M_Z^2}{M_W^2}\right)}{36 c_w^4 s_w^2}\nonumber \\
&+& \frac{120 c_w^6 +\left(48-13 s_w^2 \pi^2\right)c_w^4-6}
    {216 c_w^4 s_w^2}
+ \frac{\left(24 c_w^6+22 c_w^4+c_w^2-2\right) i \pi }
      {72 c_w^4 s_w^2} \>,
\end{eqnarray}
and the corresponding counterterms are
\begin{eqnarray}
  \label{eq:DHCT}
c_{d, h}^{(1),\, {\rm ct}} &=&
\frac{c_{p, LR}^{(1),\, {\rm ct}}}{2}
+\frac{16 c_w^4-50 c_w^2+7}{144 c_w^2 s_w^2\ \epsilon }
  \left(\frac{M_W^2}{\mu^2}\right)^{-\epsilon} \nonumber \\
&-&\frac{\left(16 c_w^4-32 c_w^2+7\right) \
\ln \left(\frac{M_Z^2}{M_W^2}\right)}{144 c_w^2 \
s_w^2}-\frac{16 c_w^4-50 \
c_w^2+7}{288 c_w^2 s_w^2}
\>.
\end{eqnarray}

\subsection{Integrals and auxiliary functions}
\label{app:integrals}

The results for the short-distance coefficients and their counterterms
have been written such that all poles in $\epsilon$ are apparent
and the remaining functions are finite. We give here their analytic
expressions.
As usual the scalar two- and three-point functions are defined by
\begin{equation}
  \label{eq:b0}
  B_0(k^2, m_1^2,m_2^2) \equiv
  \int\frac{[dl]}{(l^2-m_1^2) ((l+k)^2-m_2^2)}\>,
\qquad [dl] \equiv
  \frac{(e ^{\gamma_E} \mu^{2})^\epsilon\, d^d l}{i \pi^{d/2}}\>,
\end{equation}
and
\begin{equation}
  \label{eq:c0}
  C_0(k_1^2, k_2^2,(k_1+k_2)^2, m_1^2,m_2^2,m_3^2)  \equiv \int
  \frac{[dl]}{(l^2-m_1^2)((l+k_1)^2-m_2^2)((l+k_1+k_2)^2-m_3^2)}\,.
\end{equation}
$\partial B_0(k^2, m_1^2,m_2^2)$ is then defined as
\begin{equation}
  \label{eq:db0}
  \partial B_0(k^2, m_1^2,m_2^2) \equiv
  \frac{\partial  B_0(q^2, m_1^2,m_2^2)}{\partial q^2}\big|_{q^2=k^2}\,.
\end{equation}
The auxiliary function $\LL(k^2,m_1^2,m_2^2)$ used in the expressions
for the matching coefficients is related to the two-point function by
\begin{equation}
\label{eq:BtoLL}
B_0(k^2, m_1^2,m_2^2) = \frac{1}{\epsilon}
  \left(\frac{m_1^2}{\mu^2}\right)^{\!-\epsilon}
 + 2 -  \LL(k^2,m_1^2,m_2^2)
\end{equation}
and satisfies $\LL(k^2,m_1^2,m_2^2) = \LL(k^2,m_2^2,m_1^2) +
\ln(m_2^2/m_1^2)$. It is sufficient to give this function for the
following special arguments:
\begin{eqnarray}
\LL(0,M_W^2,M_Z^2) &=&
   1+\frac{M_Z^2}{M_W^2-M_Z^2} \ln\left(\frac{M_W^2}{M_Z^2}\right),
\nonumber \\
\LL(M_Z^2,M_W^2,M_W^2) &=&
  \frac{M_Z^2 - M_{ZW}^2}{2M_Z^2}
  \ln\left(1+\frac{M_{ZW}^2-M_Z^2}{2 M_W^2}\right)
\nonumber\\
&+&  \frac{M_Z^2 +M_{ZW}^2}{2 M_Z^2}
  \ln\left(1-\frac{M_{ZW}^2+M_Z^2}{2 M_W^2}\right), \nonumber \\
\LL(M_W^2,M_Z^2,M_W^2) &=&
  \frac{2M_W^2-M_Z^2+M_{ZW}^2}{2M_W^2}
  \ln\left(\frac{M_Z^2-M_{ZW}^2}{2 M_Z^2}\right)
\nonumber\\
&+& \frac{2M_W^2-M_Z^2-M_{ZW}^2}{2M_W^2}
  \ln\left(\frac{M_Z^2+M_{ZW}^2}{2 M_Z^2}\right),
\end{eqnarray}
where we introduced $M_{ZW}^2 \equiv \sqrt{M_Z^4-4 M_Z^2 M_W^2}$. The
explicit result for the derivative of the two-point function that is
needed reads
\begin{eqnarray}
\lefteqn{\partial B_0(M_W^2,M_W^2,M_Z^2) = } \\
&& - \frac{1}{M_W^2} \left\{ 1
+ \frac{M_W^2-M_Z^2}{2 M_W^2} \ln\left(\frac{M_Z^2}{M_W^2}\right)
+ \frac{M_Z^2 (3 M_W^2-M_Z^2)}{M_W^2 M_{ZW}^2}
  \ln\left(\frac{M_Z^2-M_{ZW}^2}{2 M_W M_Z}\right) \right\}. 
\nonumber
\end{eqnarray}
The analytic expressions of the finite three-point functions appearing
in the results given in (\ref{eq:PV0LR})--(\ref{eq:DHCT}) can all be
obtained from
\begin{eqnarray}
\label{eq:c0expl1}
\lefteqn{\hspace*{-1.5cm}C_0(0,M_W^2,-M_W^2,0,M_Z^2,M_W^2) = } &&  \\
\frac{1}{4 M_W^2}\Bigg\{ \hspace*{-0.4cm} &&
  2\, \text{Li}_2\left(1-\frac{2 M_W^2}{M_Z^2}\right)
+ 2\, \text{Li}_2\left(\frac{2 M_W^2-M_Z^2}{4 M_W^2-M_Z^2}\right)
- \text{Li}_2\left(\frac{M_Z^4}{M_{ZW}^4}\right) \nonumber \\
&-& 2\, \text{Li}_2\left(\frac{2M_W^2-M_Z^2}{M_{ZW}^2}\right)
- 2\, \text{Li}_2\left(\frac{M_Z^2-2M_W^2}{M_{ZW}^2}\right)
- \frac{\pi^2}{3} \Bigg\}\>, \nonumber\\
\lefteqn{\hspace*{-1.5cm} C_0(0,4 M_W^2,0,0,M_Z^2,M_Z^2) =} && \\
-\frac{1}{8 M_W^2}\Bigg\{\hspace*{-0.4cm}  &&
  \ln^2\left(\frac{-M_{ZW}^4}{M_Z^4}\right)
+ \ln^2\left(\frac{M_{W+Z}^2}{M_Z^2}\right)
+ 2\, \text{Li}_2\left(\frac{M_Z^4}{M_{ZW}^4}\right)
\nonumber \\
&+& 2\, \text{Li}_2\left(\frac{4 M_W^2-M_Z^2}
       {M_{W-Z}^2}\right)
+ 2\, \text{Li}_2\left(\frac{4 M_W^2-M_Z^2}
       {M_{W+Z}^2}\right)+\pi ^2
    \Bigg\}\>, \nonumber \\
\lefteqn{\hspace*{-1.5cm} C_0(-M_W^2,M_W^2,0,0,M_Z^2,M_W^2)=} && \\
-\frac{1}{2 M_W^2}\Bigg\{\hspace*{-0.4cm}  &&
   \text{Li}_2\left(-\frac{M_W^2}{M_W^2+2 M_Z^2}\right)
 - \text{Li}_2\left(\frac{M_W^2}{M_W^2+2 M_Z^2}\right)\nonumber\\
&+& \text{Li}_2\left(\frac{M_W^2}{M_W^2-M_Z^2-M_{ZW}^2}\right)
 - \text{Li}_2\left(-\frac{M_W^2}{M_W^2-M_Z^2-M_{ZW}^2}\right)\nonumber \\
&+& \text{Li}_2\left(\frac{M_W^2}{M_W^2-M_Z^2+M_{ZW}^2}\right)
- \text{Li}_2\left(-\frac{M_W^2}{M_W^2-M_Z^2+M_{ZW}^2}\right)
+ \frac{\pi^2}{4} \Bigg \}\>,  \nonumber \\
\lefteqn{\hspace*{-1.5cm} C_0(M_W^2,0,0,M_W^2,M_Z^2,0)=}  && \\
\frac{1}{M_W^2}\Bigg\{\hspace*{-0.4cm}  &&
\text{Li}_2\left(\frac{2 M_W^2}{M_Z^2+ M_{ZW}^2}\right)
+ \text{Li}_2\left(\frac{M_Z^2+M_{ZW}^2}{2M_Z^2}\right)
-\frac{\pi^2}{6}\Bigg\}\,, \nonumber \\
\lefteqn{\hspace*{-1.5cm}C_0(M_W^2,-M_W^2,0,0,0,M_Z^2)=}  &&  \\
\frac{1}{4 M_W^2}\Bigg\{\hspace*{-0.4cm}  &&
  \ln \left(\frac{2 M_W^2}{M_Z^2}+1\right)
    \left(\ln \left(\frac{2 M_W^2}{M_Z^2}+1\right)-2 i \pi \right)
- \text{Li}_2\left(\frac{M_Z^4}{\left(2M_W^2+M_Z^2\right)^2}\right)
\nonumber \\
&+& 2\, \text{Li}_2\left(1-\frac{2 M_W^2}{M_Z^2}\right)
+ 2\, \text{Li}_2\left(\frac{2 M_W^2-M_Z^2}{2 M_W^2+M_Z^2}\right)
- 2\, \text{Li}_2\left(\frac{M_Z^2-2 M_W^2}{2
   M_W^2+M_Z^2}\right)\nonumber \\
&+& 6\, \text{Li}_2\left(\frac{M_Z^2}{2 M_W^2+M_Z^2}\right)
 - \frac{2 \pi^2}{3}  \Bigg\}\>, \nonumber \\
\lefteqn{\hspace*{-1.5cm} C_0(M_W^2,0,0,0,0,M_Z^2) =}  &&  \\
\frac{1}{M_W^2} \Bigg\{\hspace*{-0.4cm} &&
\frac{1}{2} \ln ^2\left(\frac{M_W^2+M_Z^2}{M_Z^2}\right)
- i \pi  \ln \left(\frac{M_W^2+M_Z^2}{M_Z^2}\right)
+ \text{Li}_2\left(\frac{M_Z^2}{M_W^2+M_Z^2}\right)
- \frac{\pi^2}{6}\Bigg\}\>,  \nonumber
\end{eqnarray}
where we introduced $M_{W\pm Z} \equiv M_W\pm\sqrt{M_W^2-M_Z^2}$.


\end{document}